\documentclass[a4paper,11pt]{article}
\usepackage{geometry}
\usepackage{a4wide}
\usepackage{graphicx}
\usepackage{physics}
\usepackage{amsmath}
\usepackage{amssymb}

\usepackage{cite}
\usepackage{multirow,tabularx}
\usepackage{appendix}
\usepackage{tikz}
\usetikzlibrary{shapes}
\usepackage{graphicx,amsmath,amsfonts,amssymb,amsthm,euscript,braket,xcolor}
\newcommand{\be}{\begin{equation}}
\newcommand{\ee}{\end{equation}}

\newcommand{\Rmnum}[1]{\expandafter\@slowromancap\romannumeral #1@}
\newcommand{\bea}{\begin{eqnarray}}
\newcommand{\eea}{\end{eqnarray}}

\numberwithin{equation}{section}

\usepackage{subfigure}

\begin{document}
\title{\bf Analytic three-dimensional primary hair charged black holes with Coulomb-like electrodynamics and their thermodynamics}

\author{\textbf{Ayan Daripa}\thanks{ayandaripaadra@gmail.com},~~\textbf{Subhash Mahapatra}\thanks{mahapatrasub@nitrkl.ac.in}
 \\\\
 \textit{{\small Department of Physics and Astronomy, National Institute of Technology Rourkela, Rourkela - 769008, India}}
}
\date{}


\maketitle
\abstract{}
We construct and discuss new solutions of primary hair charged black holes in asymptotically Anti–de Sitter (AdS) space that have well-defined Coulomb-like potential in three dimensions.  The gauge field source to the Einstein equation is a power-Maxwell nonlinear electrodynamics with traceless energy-momentum tensor. The coupled Einstein-power-Maxwell-scalar gravity system, which carries the coupling $f(\phi)$ between the gauge and scalar fields, is analyzed, and hairy charged black hole solutions are found analytically. We consider three different profiles of the coupling functions: (i) $f(\phi)=1$, corresponding to no direct coupling between the gauge and scalar fields, (ii) $f(\phi)=e^{\phi}$, and (iii) $f(\phi)=e^{\phi^2/2}$, corresponding to their non-minimal coupling. For all these cases, the
scalar field and gauge fields, as well as curvature scalars, are regular and well-behaved everywhere outside the horizon. We further study the thermodynamics of the obtained hairy black hole in the canonical and grand-canonical ensembles and find significant changes in its thermodynamic structure due to the scalar field. In particular, for all considered coupling functions, the hairy parameter has a critical value above which the hairy black hole undergoes the Hawking/Page phase transition, whereas below which no such phase transition appears.

\section{\label{sec1}Introduction}
For many years, physicists and astronomers have been fascinated by the black hole mysteries and their fascinating nature. The study of black holes gives a unique platform to view the behaviour of matter and energy under extreme circumstances and provides insights into basic principles of general relativity. They offer a unique framework for the coexistence of strong gravity, quantum phenomena, and thermodynamics. It is now widely accepted that black holes carry both temperature and entropy and may experience phase transitions much as ordinary thermodynamic systems \cite{Hawking:1975vcx, Gibbons:1976ue, PhysRevD.7.2333}. For instance, in contrast to the Schwarzschild black hole in asymptotically flat space, black holes in anti-de Sitter (AdS) spaces are not only thermodynamically stable but also exhibit rich thermodynamic phase structures and go through phase transitions such as the Hawking/Page (black hole to thermal-AdS) or the small/large black hole phase transitions \cite{Hawking:1982dh, PhysRevD.60.064018, PhysRevD.60.104026, Cvetic:1999ne, Sahay:2010tx, Sahay:2010wi, Dey:2015ytd, Mahapatra:2016dae}.

Due to their intrinsic simplicity compared to their higher-dimensional counterparts, studies of lower-dimensional gravitational systems have drawn a lot of interest. The prime example is the three-dimensional BTZ (Banados-Teitelboim-Zanelli) black holes, which have received a lot of attention in the last three decades and have shown to be useful simplified models for investigating conceptual issues of black holes  \cite{Banados:1992wn, PhysRevD.48.1506}. For example,  general relativity becomes a topological field theory in the three dimensions, whose dynamics can be mapped holographically to the two-dimensional conformal field theory (CFT) living at the boundary of spacetime \cite{Brown:1986nw}. The BTZ black holes, therefore
provide a natural arena to test the deep and fundamental
principles of gauge/gravity duality \cite{Maldacena:1997re}. The use of conformal boundary conserved charges and symmetric algebra to calculate the entropy of BTZ black holes provides a prime example of the usefulness of lower-dimensional gravity systems to get possible insight into quantum gravity \cite{Strominger:1997eq}. In addition, the Chern-Simons (CS) formulation of three-dimensional gravity models has made them quintessential for investigating general properties of gravity, and in particular its relationship with gauge field theories \cite{Achucarro:1986uwr, Witten:1988hc}. Indeed, despite having various contrasting features compared to their higher-dimensional counterparts -- such as not containing any curvature singularity or being locally equivalent to
pure $AdS_3$ -- the BTZ black holes do exhibit many of
their fundamental features, such as the presence of event
and Cauchy horizons, or their thermodynamic and holographic interpretations \cite{Carlip:1995qv}. For these reasons, the lower-dimensional models continue to be the focus of considerable interest in gravitational theories \cite{Luongo:2023xaw,Sajadi:2023ybm}.

The repertory of three-dimensional black hole solutions has greatly broadened in a number of ways since the seminal discovery of the BTZ solution. In addition to the traditional Maxwell term \cite{Martinez:1999qi}, higher-order curvature terms \cite{Bergshoeff:2009hq}, higher-rank tensor fields \cite{Perez:2013xi}, gravitational Chern-Simons terms \cite{Ammon:2011nk} etc have been added in the gravity action. The extension also includes a variety of other matter sources.  These advancements have improved our knowledge of lower-dimensional black hole solutions and their interactions with various types of matter. The Maxwell Lagrangian, $\mathcal{L}(\mathcal{F})=\mathcal{F}=F_{\mu\nu}F^{\mu\nu}$, in particular, leads to an interesting but undesirable behaviour in three dimensions, i.e., the electric field and potential of the black hole are now proportional to $1/r$ and $\log{r}$ respectively, $r$ being the radial coordinate, as opposed to $1/r^2$ and $1/r$ behaviour in four dimensions. Consequently, divergence terms appear in the metric and gauge field, as $\log(r)$ blows up at $r=0$ and $r=\infty$, making the charged BTZ solution unattractive. This raises the potential question: is it possible to construct a Lagrangian for the gauge field that will give us a gauge field solution just like in the case of $(3+1)$ dimensions, i.e., a regular and well-behaved gauge field solution? This question was addressed in \cite{Cataldo:2000we}, where the answer was found in the affirmative. The essential idea was to require, as in four dimensions, the trace of the gauge field energy-momentum tensor to be zero. This condition uniquely constrained the gauge field Lagrangian to be power-like with exponent $3/4$, i.e., for the power law Lagrangian $\mathcal{L}(\mathcal{F})=(s\mathcal{F})^{p}$, the trace $$T=T_{\mu\nu}g^{\mu\nu}=3\mathcal{L}(\mathcal{F})-4\mathcal{F}\mathcal{L},_{\mathcal{F}}= (\frac{3}{2}-2p)(s\mathcal{F})^p$$ vanishes for the exponent value $p=3/4$. It is straightforward to check that this traceless nonlinear electrodynamics \big[$\mathcal{L}(\mathcal{F})=(s\mathcal{F})^{3/4}$\big] causes the radial dependence of electric field and potential to be $1/r^2$ and $1/r$ in $(2+1)$ dimensions, respectively. Moreover, this energy-momentum tensor also fulfils the weak energy conditions. Adding this non-linear Lagrangian to the Einstein-Hilbert action further gives three-dimensional black hole solutions with finite gauge field everywhere \cite{Cataldo:2000we}, unlike the usual charged BTZ black hole \cite{Martinez:1999qi}. Subsequently, several works have investigated various properties of these $(2+1)$-dimensional black holes with a Coulomb-like field \cite{Garcia-Diaz:2017cpv,Larranaga:2008qw,Larranaga:2008fn,Balart:2009et,Mazharimousavi:2011nd,Balart:2019uok,Rincon:2018dsq,Rincon:2017goj,Dehghani:2016agl,Cataldo:2020cxm,HabibMazharimousavi:2013duq,Gonzalez:2021vwp,Amirabi:2021uam,Aragon:2021ogo,Sheykhi:2012zz}.

In a similar vein, numerous hairy black hole solutions involving self-interacting real scalar field, both minimally and non-minimally coupled, have been investigated in three-dimensional spacetime following the initial work of \cite{Martinez:1996gn, Henneaux:2002wm}. The investigation of the interplay between scalar fields and black holes in three dimensions has not only improved our understanding of the interaction between them but also has opened up a wide range of their potential applications, especially in the context of AdS spaces, i.e., the scalar-gravity models are very pertinent to holography and are useful resources for studying lower-dimensional condensed matter systems at strong couplings. The desirable trait of analytical tractability of three-dimensional scalar gravity models has also greatly piqued our curiosity \cite{Chan:1994qa,Chan:1996rd,Ayon-Beato:2004nzi,Banados:2005hm,Correa:2010hf,Correa:2012rc,Xu:2013nia,Xu:2014uha,Cardenas:2014kaa,Tang:2019jkn,Dehghani:2017thu,Dehghani:2017zkm,Bueno:2021krl,Ahn:2015uza,
Karakasis:2022fep,Zou:2014gla,Sadeghi:2013gmf,Zhao:2013isa,Bravo-Gaete:2014haa,Baake:2020tgk,Bravo-Gaete:2020ftn,Karakasis:2023ljt}. Nonetheless, it should be emphasised that not all scalar field-dressed black hole solutions obtained in three dimensions have desirable physical properties. In many cases, the scalar field not only shows logarithmic radial dependency, thereby making them unsatisfactory but also the geometry does not asymptote to AdS at the boundary  \cite{Dehghani:2017thu,Karakasis:2022fep}.

Another important reason for greatly investigating scalar-gravity systems is due to their connections with the no-hair theorem. The no-hair theorem basically asserts that black holes in the asymptotically flat spaces are uniquely described by their mass, charge, and angular momentum \cite{Ruffini:1971bza}. Many  strong arguments and critical remarks in favour of the
no-scalar hair theorem for asymptotic flat space-time were
discussed in \cite{Bekenstein:1971hc, Sudarsky:1995zg, Heusler:1992ss}, and has long been thought to apply to black holes in general. For a review on the issue of scalar hair in asymptotic flat spaces, see \cite{Herdeiro:2015waa}. While multiple subsequent investigations have endorsed the original no-hair theorem for black holes \cite{Israel:1967wq, Wald:1971iw, Carter:1971zc, Robinson:1975bv, Mazur:1982db, Mazur:1984wz, Teitelboim:1972qx, Volkov:1990sva}, it is crucial to keep in mind that it is not a theorem in the strict mathematical sense. Many counterexamples have contested the no-hair theorem over time. The Einstein-Yang-Mills theory \cite{Bizon:1990sr, Volkov:1989fi, Greene:1992fw}, dilatonic black holes \cite{Kanti:1995vq}, black holes with Skyrme hairs \cite{Luckock:1986tr, Droz:1991cx}, black hole hair with tensor vacuum \cite{Ovalle:2020kpd, Mahapatra:2022xea} are a few instances of these counterexamples. In recent years, many works addressing various physical properties of the hairy black holes in different asymptotic spaces have appeared; for a necessarily biased selection, see \cite{Torii:1998ir,Torii:2001pg,Winstanley:2002jt,Martinez:2004nb,Martinez:2006an,Hertog:2004dr,Henneaux:2006hk,Herdeiro:2014goa,Herdeiro:2018wub,Hertog:2006rr,Kolyvaris:2010yyf,Gonzalez:2013aca,Dias:2011tj,
Bhattacharyya:2010yg,Dias:2011at,Anabalon:2012ih,Kolyvaris:2011fk,Kolyvaris:2013zfa,Ballon-Bayona:2020xls,Guo:2021ere,Anabalon:2013qua,Astefanesei:2019ehu,
Priyadarshinee:2021rch,Mahapatra:2020wym,Guo:2023mda,Kiorpelidi:2023jjw,Theodosopoulos:2023ice}.

The analytical tractability of scalar gravity systems in three dimensions have made them great laboratories for discussing the no-hair theorem \cite{Garcia-Diaz:2017cpv}. It is then natural to construct and investigate analytic scalar hair black hole solutions with a well-behaved Coulomb-like structure for the gauge field using the Einstein-power Maxwell-scalar gravity system. Such solutions have been obtained in relatively few systems \cite{Dehghani:2022aae,Dehghani:2018hpb, Dehghani:2019dab,Hendi:2017mgb}. However, the scalar field in these hairy solutions depends logarithmically on the radial coordinate and therefore, diverges at the boundary. Moreover, the solutions do not asymptote to AdS at the boundary, thereby severely restricting their potential applicability.

From the thermodynamic perspective as well, the scalar hair can change the thermodynamic phase structure of three-dimensional black holes considerably. In particular, the black hole temperature exhibits multiple branches in four and higher dimensions, allowing for the possibility of phase transitions as the temperature varies. Two prominent examples of such phase transitions are the Hawking/Page phase transition between AdS black holes and thermal-AdS in the grand canonical ensemble and the liquid/gas type phase transition between small and large black holes in the fixed charge ensemble. Whereas the temperature profile in BTZ black hole exhibits only one branch, displaying no phase transitions in both charged and uncharged cases. However, recently, it was observed that certain three-dimensional hairy black holes, obtained from the potential reconstruction technique, can exhibit such phase transitions in the presence of nontrivial primary scalar hair \cite{Priyadarshinee:2023cmi}. In particular, it was observed that depending on the coupling function $f(\phi)$ between the gauge and scalar field, the primary scalar hair can greatly influence the phase structure of three-dimensional black holes and resemble them to that of higher-dimensional charged AdS black holes.

Since the addition of scalar and gauge fields to Einstein's gravity usually generates unexpected and exciting features in black hole solutions, it is instructive to find new exact solutions to the Einstein-power Maxwell-scalar gravity system for arbitrary coupling functions and investigate how the geometrical and thermodynamical properties of black holes are altered in the presence of a scalar field. In particular, it is interesting and desirable to have three-dimensional primary hair black hole solutions, with not just a regular profile of the scalar field but also of the gauge field, as such solutions might have applications in applied holography. In this work, we find a family of such solutions.

This paper introduces analytical charged primary hair black hole solutions in three dimensions with a power Maxwell field, whose thermodynamic structure is somewhat similar to that of charged AdS black holes in higher dimensions. We focus on the Einstein-power Maxwell-scalar gravity system, which has a coupling function $f(\phi)$ between the scalar and power Maxwell fields, and use the potential reconstruction technique \cite{ Dudal:2017max, Dudal:2021jav, Bohra:2020qom, Mahapatra:2018gig, Bohra:2019ebj, He:2013qq, Arefeva:2018hyo,Arefeva:2022avn,Arefeva:2020byn, Alanen:2009xs} to simultaneously solve the coupled Einstein-power Maxwell-scalar field equations in terms of functions $f(\phi)$ and $A(z)$ (see the next section for details). The different forms of $f(\phi)$ and $A(z)$ then allow us to construct a different family of hairy black hole solutions. To make the analysis and findings more thorough, we choose three different but physically motivated forms of the coupling function: (i) $f(\phi)=1$, (ii) $f(\phi)=e^{\phi}$, and (iii) $f(\phi)=e^{\phi^2/2}$. While the second and third coupling functions relate to a non-minimal coupling between the scalar and gauge fields, the first coupling function indicates that there is no direct coupling between them. The primary reason for taking into account such coupling functions is the fact that they have recently been thoroughly investigated in a variety of higher-dimensional hairy black hole contexts, from scalarization to holographic model construction \cite{Herdeiro:2018wub}, and have consistently contributed to our understanding of the hairy aspects of black holes. Therefore, it is intriguing to look into how these coupling functions affect the geometrical and thermodynamical features of three-dimensional black holes as well. We similarly take a particularly simple form of $A(z)=-a^2z^2$, which enables us to introduce the parameter $a$, which regulates the strength of the scalar field.

For all forms of $A(z)$ and $f(\phi)$ considered here, the found hairy black hole solutions exhibit many attractive features. This includes: (i) the scalar field is finite and well-behaved everywhere in the outer horizon region and falls off at the asymptotic AdS boundary; (ii) the gauge field is also finite everywhere outside the horizon; (iii) the curvature scalars, such as the Kretschmann and Ricci scalars, are also finite everywhere outside the horizon, suggesting no additional singularity in the hairy solution than those already present in the nonhairy case; (iv) these hairy solutions can be analytically continued to standard BTZ solution in the limits $\{a\rightarrow 0, q_e\rightarrow 0\}$; and (v) the potential is bounded from above from its UV boundary value, thereby satisfying the Gubser criterion to have a well-defined boundary theory \cite{Gubser:2000nd}.

We then analyze the thermodynamic structure of the obtained hairy solutions in the canonical and grand-canonical ensembles and find that it changes significantly when the hairy parameter $a$ is turned on. In particular, for all considered coupling functions, and in both canonical and grand-canonical ensembles, a critical value of the hairy parameter $a=a_c$ appears (which is a $f(\phi)$ dependent quantity) above which the hairy black hole undergoes the Hawking/Page phase transition to thermal-AdS phase, whereas no such phase transition appears below $a_c$. For the uncharged $q_e=0$ case, the Hawking/Page phase transition exists for all finite values of $a$. The corresponding transition temperature also increases monotonically with $a$. Moreover, the critical value $a_c$ turns out to be a $\mu_e$ and $q_e$ dependent quantity in the grand-canonical and canonical ensembles respectively, i.e., its magnitude increases as $\mu_e$ or $q_e$ increases. This thermodynamic behavior of the hairy black hole is therefore analogous to the BTZ black hole for $a<a_c$ whereas it resembles RN-AdS black hole in the grand-canonical ensemble for $a>a_c$. Moreover, we find that these primary hair black holes are also thermodynamically stable as they exhibit positive specific heat.

The paper is structured as follows: In Sec.~2, we discuss the three-dimensional Einstein-power Maxwell-scalar gravity model and present its analytic solution in terms of two functions $f(\phi)$ and $A(z)$. In Sec.~3, we study the geometrical and thermodynamical properties of hairy black hole solution for the coupling $f(\phi)=1$.  In Sec.~4 and 5, we repeat the calculations with different couplings $f(\phi)=e^{\phi}$, and $f(\phi)=e^{\phi^2/2}$. Finally, in Sec.~6, we conclude and summarize our results.

\section{\label{sec2}Hairy charged black hole Solution}
To construct hairy charged black holes with power Maxwell-like electrodynamics in three dimensions, we start with the following gravity action
\begin{equation}
{S}=-\frac{1}{16\pi G_{3}}\int \,d^{3}x \sqrt{-g}\left[R-\frac{1}{2}g^{\mu\nu}\partial_{\mu}\phi\partial_{\nu}\phi-V({\phi})+\frac{f(\phi)}{4}(s\mathcal{F})^{\frac{3}{4}} \right]\,,
\label{emsaction}
\end{equation}
where $R$ denotes the Ricci scalar of the three-dimensional manifold $\mathcal{M}$, $V(\phi)$ is the potential of the scalar field $\phi$, $f(\phi)$ corresponds to the coupling between the gauge and scalar fields, and $F_{\mu\nu}$ is the field strength tensor. In terms of the four-potential $B_{\mu}$, $F_{\mu\nu}$ is expressed as $F_{\mu\nu}$= $\partial_{\mu}B_{\nu}-\partial_{\nu}B_{\mu}$. Note that, as discussed in the introduction, the electromagnetic part of the action is traceless.

The variation of Eq.~(\ref{emsaction}) gives the following Einstein, gauge, and scalar field equations:
\begin{eqnarray}
& & R_{MN} - \frac{1}{2}g_{MN} R + \frac{1}{2} \biggl(\frac{g_{MN}}{2} \partial_{P}\phi \partial^{P}\phi -\partial_{M}\phi \partial_{N}\phi  + g_{MN} V(\phi)  \biggr) \nonumber \\
& & - \frac{f(\phi)}{4} (s\mathcal{F})^{-\frac{1}{4}} \biggl(-\frac{g_{MN}}{2} (s\mathcal{F}) + \frac{3}{2} (s F_{MP}F_{N}^{\ P})\biggr)=0 \,,
\label{einsteineqn}
\end{eqnarray}
\begin{equation}
\partial_\mu[\sqrt{-g}f(\phi)(s\mathcal{F})^{-\frac{1}{4}}F^{\mu\nu}]=0\,,
\label{maxwelleqn}
\end{equation}
\begin{equation}
\frac{1}{\sqrt{-g}}\partial_\mu[\sqrt{-g}\partial^{\mu}\phi]+ \frac{1}{4}f'(\phi)(s\mathcal{F})^{\frac{3}{4}}=\frac{\partial V(\phi)}{\partial \phi}\,,
\label{scalarfieldeqn}
\end{equation}
where $f'(\phi)$ denotes the derivative of the coupling function with respect to the field $\phi$. We consider the following Ans$\ddot{a}$tze for the metric, scalar field, and gauge field to construct a static and spherically symmetric ($S^1$) hairy black hole solution in three dimensions:
\begin{eqnarray}
& & ds^2=\frac{L^2}{z^2}\biggl[-g(z)dt^2 + \frac{ e^{2A(z)} dz^2}{g(z)} + d\theta^2 \biggr]\,, \nonumber \\
& & \phi=\phi(z), \ \ B_{\mu}=B_{t}(z)\delta_{\mu}^{t} \,,
\label{metricansatz}
\end{eqnarray}
where $A(z)$ is the form factor, whose form will be crucial in determining the hairy black hole solution and the corresponding thermodynamics, $L$ is the AdS length scale, which we will set to one from here on for simplicity, and $g(z)$ is the blackening function. The radial coordinate $z$ ranges from $z = 0$ (asymptotic boundary) to $z= z_{h}$ (black hole horizon radius) or to $z =\infty$ for thermal-AdS (without horizon).

There is only one non-zero component of Faraday's tensor in the geometry defined by \eqref{metricansatz}, and that is $F_{tz}$=$-B'_{t}(z)$. So we can write $\mathcal{F}=2F^2_{tz}g^{tt}g^{zz}=2B'^{2}_{t}(z)g^{tt}g^{zz}$. As a result, we set $s=-1$ to have real solutions for the gauge field while considering the electromagnetic Lagrangian's fractional power and maintaining generality. Now using Eq.~(\ref{maxwelleqn}), we get
\begin{equation}
F_{tz}=-B'_{t}(z)= -\frac{\sqrt{2} q_{e} e^{A(z)}}{f(\phi)^2}\,,
\label{electricfield}
\end{equation}
where $q_e$ is an integration constant related to the charge of the black hole (see below). Similarly, by substituting (\ref{metricansatz}) into (\ref{einsteineqn}), we get the following three Einstein equations of motion,
\begin{eqnarray}
tt\equiv g'(z)-g(z)\left(\frac{2}{z}+2A'(z)+\frac{1}{2} z \phi '(z)^2\right)-\frac{V(\phi) e^{2A(z)}}{z}-\frac{f(\phi)e^{2A(z)}(s\mathcal{F})^{\frac{3}{4}}}{8z}=0\,,
\label{ttcomponent}
\end{eqnarray}
\begin{eqnarray}
zz\equiv g'(z)-g(z)\left(\frac{2}{z}-\frac{1}{2} z \phi '(z)^2 \right)-\frac{V(\phi)e^{2A(z)}}{z}-\frac{f(\phi)e^{2A(z)}(s\mathcal{F})^{\frac{3}{4}}}{8z}=0 \,,
\label{zzcomponent}
\end{eqnarray}
\begin{eqnarray}
\theta\theta\equiv g''(z) - g'(z)\left(\frac{2}{z}+A'(z)\right) + g(z)\left(\frac{2}{z^2}+\frac{2A'(z)}{z} + \frac{1}{2} \phi '(z)^2 \right) + \frac{V(\phi)e^{2A(z)}}{z^2} \nonumber\\
-\frac{f(\phi)e^{2A(z)}(s\mathcal{F})^{\frac{3}{4}}}{4z^2}=0 \,.
\label{thetathetacomponent}
\end{eqnarray}
The above three equations can be further rearranged into the following equations,
\begin{equation}
g''(z)-g'(z)\left(\frac{1}{z}+A'(z) \right)-\frac{3f(\phi)e^{2A(z)}(s\mathcal{F})^{\frac{3}{4}}}{8z^2}=0\,,
\label{g(z)eqn}
\end{equation}
\begin{equation}
\frac{\phi'(z)^2}{2}+\frac{A'(z)}{z}=0\,,
\label{phi(z)eqn}
\end{equation}
\begin{eqnarray}
& & g''(z)- g'(z)\left(\frac{3}{z}+A'(z) \right)+ g(z)\left(\frac{4}{z^2}+\frac{2A'(z)}{z} \right) \nonumber\\
& & \frac{2V(\phi) e^{2A(z)}}{z^2}-\frac{f(\phi)e^{2A(z)}(s\mathcal{F})^{\frac{3}{4}}}{8z^2}=0\,.
\label{v(phi)eqn}
\end{eqnarray}
Similarly, the scalar field equation of motion is given by
\begin{eqnarray}
\phi ''(z) - \phi '(z) \left(A'(z)-\frac{g'(z)}{g(z)}+\frac{1}{z}\right)
   -\frac{e^{2 A(z)}}{z^2 g(z)} \frac{\partial V(\phi)}{\partial \phi} -\frac{e^{2 A(z)} (s\mathcal{F})^{3/4}}{4 z^2 g(z)} \frac{\partial f(\phi)}{\partial \phi}   =0\,.
\label{scalarEOM}
\end{eqnarray}
Therefore, overall, we have five equations of motion in the gravity system of Eq.~(\ref{emsaction}). However, only four of them are independent. It can be explicitly checked that the last Eq.~(\ref{scalarEOM}) follows from the Bianchi identity and is therefore redundant. Below we will choose Eq.~(\ref{scalarEOM}) as a constrained equation and consider the rest of the equations as independent. We now impose the following boundary conditions to solve these equations:
\begin{eqnarray}
&& g(0)=1, \ \ \text{and} \ \ g(z_h)=0, \nonumber \\
&& B_{t}(0)= \mu_e, \ \ \text{and} \ \  B_{t}(z_h)=0, \nonumber \\
&& A(0) = 0 \,.
\label{boundarycdt2}
\end{eqnarray}
The boundary conditions at $z=0$ are chosen to ensure that the spacetime asymptotes to AdS at the boundary. The parameter $\mu_e$ is the leading term of the near boundary expansion of the gauge field $B_t(z)$ and corresponds to the chemical potential of the theory. Using Gauss's theorem, we can also find a relation between $\mu_e$ and the electric charge of the black hole (see the discussion below). In addition to these boundary conditions, we further require that the scalar field goes to zero at the boundary $\phi(0) = 0$ and must remain real throughout the bulk.

Using the above boundary conditions and solving Eq.~(\ref{electricfield}), we get the following solution for the gauge field
\begin{eqnarray}
B_{t} (z) = \sqrt{2}q_{e} \int_{z}^{z_h} \, d\xi~ \frac{e^{A(\xi)}}{f^2(\xi)} \,.
\label{Atsol}
\end{eqnarray}
Similarly, by solving Eq.~(\ref{g(z)eqn}), the solution for $g(z)$ is
\begin{eqnarray}
& & g(z) =  C_1 + \int_0^z \, d\xi \ e^{A(\xi)} \xi \biggl[ C_{2} + \mathcal{K}(\xi) \biggr] \,,
\label{gsol}
\end{eqnarray}
where
\begin{eqnarray}
& & \mathcal{K}(\xi)= \frac{3 q_{e}^{3/2}}{2\sqrt{2}} \int \, d\xi~ \frac{e^{A(\xi)}}{f^2(\xi)}  \,,
\label{gsol}
\end{eqnarray}
where the integration constants $C_1$ and $C_2$ are,
\begin{eqnarray}
C_1 = 1,  \ \ \ \ \ C_2 =- \frac{1+ \int_0^{z_h} \, d\xi~ e^{A(\xi)} \xi \mathcal{K}(\xi) }{ \int_0^{z_h} \, d\xi~ e^{A(\xi)} \xi}  \,.
\end{eqnarray}
The expression of scalar field $\phi$ can be similarly found by solving Eq.~(\ref{phi(z)eqn})
\begin{eqnarray}
\phi(z) = \int \, dz \ \sqrt{\frac{ - 2 A'(z)}{z}} + C_{3} \,,
\label{phisol}
\end{eqnarray}
where $C_{3}$ can be obtained by demanding $\phi$ to vanishes near the asymptotic boundary, i.e., $\phi |_{z=0}\rightarrow 0$. Lastly, the potential $V$ can be found from Eq.~(\ref{v(phi)eqn}),
\begin{eqnarray}
V(z) & = &  \frac{1}{16} e^{-2 A(z)} \left(8 z^2 A'(z)+24 z\right) g'(z)+\frac{1}{16}
   e^{-2 A(z)} g(z) \left(-16 z A'(z)-32\right) \nonumber\\
& &   +\frac{z^3 f(z) \left(e^{-2
   A(z)} B_{t}'(z)^2\right)^{3/4}}{8 \sqrt[4]{2}}-\frac{1}{2} z^2 e^{-2
   A(z)} g''(z) \,.
\label{Vsol}
\end{eqnarray}
It is thus clear that Eqs.~(\ref{Atsol})-(\ref{Vsol}) exhibit a closed-form analytic solution of the gravity system of Eq.~(\ref{emsaction}) in $(2+1)$-dimensions in terms of two functions $A(z)$ and $f(z)$. The constructed hairy solution will depend only on $A(z)$ once the coupling function $f(\phi)$ is fixed. However, different forms of $A(z)$ and $f(\phi)$ will correspond to different $V(z)$, i.e., various $A(z)$ and $f(\phi)$ will ascribe to different $(2+1)$-dimensional hairy black hole solutions. Therefore, by selecting different forms of $A(z)$ and $f(\phi)$, one may systematically construct a vast family of physically permissible primary hair charged black hole solutions for the Einstein-power Maxwell-scalar gravity system in $(2+1)$-dimensions.

In the context of applied gauge/gravity duality, the forms of $A(z)$ and $f(\phi)$ are often determined or fixed by demanding a sensible dual boundary field theory. In particular, suitable forms of $A(z)$ and $f(\phi)$ are typically taken depending on the sort of boundary field theory one is interested in. For instance, in the field of holographic QCD, the forms of these functions are typically fixed by requiring the dual boundary field theory to exhibit genuine QCD characteristics, such as confinement/deconfinement phase transition \cite{Jain:2022hxl,Shukla:2023pbp,Jena:2022nzw}, confinement in the quark sector, linear Regge trajectory for the excited meson mass spectrum, etc. In such model-building cases, the form $A(z)=-a^2z^2$ is generally considered \cite{Dudal:2017max,Bohra:2019ebj,Arefeva:2018hyo}.

However, we can also take a more liberal and phenomenological approach and investigate various physically motivated forms of $A(z)$ and $f(\phi)$ to thoroughly discuss the effects of scalar hair and make qualitative arguments about the stability and thermodynamics of the hairy charged black holes in three dimensions with Maxwell-like electrodynamics, without worrying too much about the dual boundary field theory. Here, we take such an approach. In particular, we take three physically motivated forms of the coupling function: (i) $f(\phi)=1$; (ii) $f(\phi)=e^{\phi}$; and (iii) $f(\phi) = e^{\phi^2/2}$. As mentioned earlier, these three types of couplings have recently received a lot of attention in several contexts involving hairy black holes, for example, see \cite{Herdeiro:2018wub}. It is, therefore, interesting to examine how these coupling functions influence the hairy black hole structure in three dimensions as well. Similarly, following \cite{Priyadarshinee:2021rch}, we focus on a particularly straightforward form of $A(z) = -a^2z^2$. In addition to being simpler, this form of $A(z)$ is particularly chosen as it gives us more control over the integrals that show up in the solutions of various geometric functions discussed above. This form of $A(z)$ has also been widely employed in the holographic QCD literature; for instance, see \cite{Bohra:2019ebj,Dudal:2017max}.  With the considered form of $A(z)=-a^2 z^2$, the parameter $a$ determines the strength and backreaction of the scalar field. As a result, the scalar field back-reaction drops to zero when the parameter $a$ goes to zero. Thus, as desired, one returns to the charged BTZ black hole-like solution with Maxwell-like electrodynamics in the limit $a\rightarrow 0$.

There are also other important reasons for taking the above-mentioned forms of $f(\phi)$ and $A(z)$:
\begin{itemize}
\item They ensure that the obtained hairy solution asymptotes to AdS, i.e., at the  boundary $z\rightarrow 0$, we have
\begin{eqnarray}
& & V(z)|_{z\rightarrow 0} = -\frac{2}{L^2} + \frac{m^2\phi^2}{2}+\dots \,, \nonumber\\
 & &  V(z)|_{z\rightarrow 0} =  2\Lambda + \frac{m^2\phi^2}{2}+\dots   \,,
\label{Vsolexp}
\end{eqnarray}
where, as usual, $\Lambda=-\frac{1}{L^2}$ is the negative cosmological constant in three dimensions. Similarly, the Ricci scalar $R$ approaches $-6/L^2$ asymptotically. This, together with the fact that $g(z)|_{z\rightarrow 0}=1$, indeed ensures that the obtained solutions asymptote to AdS at the boundary. Moreover, the mass of the scalar field $m^2=-1$ also satisfies the Breitenlohner-Freedman bound for stability in AdS space, i.e., $m^2\geq-1$ \cite{Breitenlohner:1982jf}.

\item  Furthermore, as we will show shortly, the hairy solutions satisfy the Gubser criterion to have a well-defined dual boundary field theory \cite{Gubser:2000nd}.

\item These forms of $f(z)$ and $A(z)$ also ensure that the null energy condition is always respected in our gravity model. The null energy condition can be expressed as
\begin{eqnarray}
T_{MN}\mathcal{N}^M \mathcal{N}^N \geqslant 0 \,,
\label{NEC}
\end{eqnarray}
where the null vector $\mathcal{N}^{M}$ satisfies the condition $g_{MN}\mathcal{N}^M \mathcal{N}^N=0$ and $T_{MN}$ is the energy-momentum tensor of the matter fields. The null vector $\mathcal{N}^{M}$ can be chosen as
\begin{eqnarray}
\mathcal{N}^M= \frac{1}{\sqrt{g(z)}}\mathcal{N}^{t} + \frac{\cos{\alpha}\sqrt{g(z)}}{e^{A(z)}} \mathcal{N}^{z} + \sin{\alpha} \mathcal{N}^{\theta}  \,,
\label{nullvector}
\end{eqnarray}
for arbitrary parameter $\alpha$. The NEC then becomes
\begin{eqnarray}
T_{MN}\mathcal{N}^M \mathcal{N}^N  =  \frac{3 z f(z) \sin ^2\alpha \left(e^{-2 A(z)}
   B_{t}'(z)^2\right)^{3/4}}{8 \sqrt[4]{2}}+\frac{1}{2} e^{-2 A(z)} g(z)
   \cos ^2\alpha \phi '(z)^2 \geqslant 0  \,,\nonumber\\
\label{nullvector}
\end{eqnarray}
which is always satisfied everywhere outside the horizon for the chosen forms of $A(z)$ and $f(\phi)$.
\end{itemize}
Now that the hairy black hole solutions have been constructed, let us write down the expressions for various thermodynamic quantities.  This will be useful later in discussing hairy black hole thermodynamics. The black hole temperature ($T$) and entropy ($S_{BH}$) are given by:
\begin{equation}
\begin{aligned}
T=&-\frac{e^{-A(z_{h})}g'(z_{h})}{4\pi},\\
S_{BH}=& \frac{\mathcal{A}}{4G_{3}}=\frac{2\pi}{4G_{3}z_{h}},
\label{tempandentropy}
\end{aligned}
\end{equation}
where $\mathcal{A}=2\pi/z_h$ is the area of the event horizon. Similarly, we can compute the electric charge $Q$ of the black hole by measuring the flux of the electric field at the boundary,
\begin{equation}
Q=\frac{1}{16\pi G_{3}}\int f(\phi) \left(s \mathcal{F} \right)^{-\frac{1}{4}} F_{\mu\nu}u^{\mu}n^{\nu}d\theta \,,
\label{electriccharge}
\end{equation}
where $u^{\mu}$ and $n^{\nu}$ are the unit space-like and time-like normals to the constant radial surface, respectively
\begin{equation}
\begin{aligned}
u^{\mu}=&\frac{1}{\sqrt{-g_{tt}}}\delta^{\mu}_{t}=\frac{z}{L\sqrt{g(z)}}\delta^{\mu}_{t}\,,\\
n^{\nu}=&\frac{1}{\sqrt{g_{zz}}}\delta^{\nu}_{z}=\frac{z\sqrt{g(z)}}{Le^{A(z)}} \delta^{\nu}_{z}\,,
\label{normals}
\end{aligned}
\end{equation}
and d$\theta$ represents the integration across the one-dimensional boundary space. Using \eqref{electricfield}, and after simplification, we obtain the following expression of the black hole charge:
\begin{equation}
Q=\frac{\sqrt{q_e}}{16\pi G_{3}}\,.
\label{simplifiedcharge}
\end{equation}
We can also find a relation between the electric charge and the corresponding conjugate chemical potential $\mu_e$. The chemical potential is the leading term of the near boundary expansion of the gauge field $B_t(z)$. Substituting $B_t(z)$ from Eq.~(\ref{Atsol}) into Eq.~(\ref{electricfield}), we get
\begin{equation}
\mu_e=\sqrt{2} q_e \int_{0}^{z_h} \, d\xi~ \frac{e^{A(\xi)}}{f^2(\xi)} \,.
\label{electricpotential}
\end{equation}
With hairy black hole solution in hand, let us also mention that there exists another solution to the gravity equations of motion. This solution does not exhibit the horizon and is called thermal-AdS.
 \footnote{We will refer to this without horizon solution as thermal-AdS here for convenience, even though this solution does not have a constant curvature throughout the spacetime.} The thermal-AdS solution can be derived from the black hole solution by taking the limit $z_{h}\to\infty$. Depending on the nature of $A(z)$, the thermal-AdS may have a non-trivial structure in the bulk. However, due to the imposed boundary conditions (\ref{boundarycdt2}), and just like in the case of black hole solution, it always asymptotes to AdS at the boundary. Intriguingly, as we shall show later, depending on the magnitudes of $a$ and $\{\mu_e, q_e\}$, there may also be a Hawking/Page type thermal-AdS/black hole phase transition between these two solutions.

\section{\label{sec:level1} Hairy Black hole solution with coupling $f(\phi)$ = 1 }
In this section, we will first look at the geometric and thermodynamic properties of the hairy black hole solution for the simplest coupling function $f(\phi)=1$. With the considered form factor $A(z)=-a^2z^2$, the solution for the scalar field is:
\begin{equation}
\phi(z)=2az \,.
\label{phivalue1}
\end{equation}
From the equation \eqref{electricfield}, we have $F_{tz}=-\sqrt{2}q_e e^{-a^2z^2}$. This gives us the gauge field solution
\begin{equation}
B_{t}(z)= \frac{\sqrt{\pi}q_e \left(\text{erf}(a z_h)-\text{erf}(a z)\right)}{\sqrt{2}a} \,,
\label{gaugefield1}
\end{equation}
where \text{erf} is the error function. Notice that the scalar field vanishes in the limit $a\to0$. Similarly, in the limit $a\to0$, the gauge field also reduces to
\begin{equation}
B_{t}(z)= \sqrt{2} q_e (z_h-z)\,,
\end{equation}
indicating that the electrodynamics employed here in three dimensions for a nonhairy black hole is Maxwellian type. Using Eq.~(\ref{electricpotential}), we can further find the relation between $\mu_e$ and $q_e$
\begin{equation}
q_e = \sqrt{\frac{2}{\pi}} \frac{a \mu_e}{\text{erf}(a z_h)}\,.
\end{equation}
Now, Using the equation \eqref{g(z)eqn}, we have the following solution for $g(z)$:
\begin{eqnarray}
 g(z) & = & \frac{1-e^{a^2 \left(z_h^2-z^2\right)}}{1-e^{a^2 z_h^2}} + \frac{3 \sqrt{\pi } e^{-a^2 z^2} q_e^{3/2} \left(e^{a^2 z^2}
   \text{erf}\left(\sqrt{2} a z\right)-\sqrt{2} \text{erf}(a z)\right)}{16
   a^3} \nonumber\\
& &   + \frac{3 \sqrt{\pi } e^{-a^2 z^2} \left(e^{a^2 z^2}-1\right) q_e^{3/2}
   \left(\sqrt{2} \text{erf}\left(a z_h\right)-e^{a^2 z_h^2}
   \text{erf}\left(\sqrt{2} a z_h\right)\right)}{16 a^3 \left(e^{a^2
   z_h^2}-1\right)}\,.
\label{g(z)1}
\end{eqnarray}
Note that in the limit $a\to0$, this expression reduces to the charged black hole expression found in \cite{Cataldo:2000we} with a Maxwell-like potential, i.e.,
\begin{eqnarray}
g(z)=1-\frac{z^2}{z_h^2}-\frac{z^2  z_h q_e^{3/2}}{2 \sqrt{2}}+\frac{z^3 q_e^{3/2}}{2
   \sqrt{2}}\,.
   \label{gza0limit}
\end{eqnarray}
Similarly, we have calculated $V(z)$, but since it is rather long and not very informative, we prefer not to write it down here for brevity.

\begin{figure}[ht]
\subfigure[]{
\includegraphics[scale=0.4]{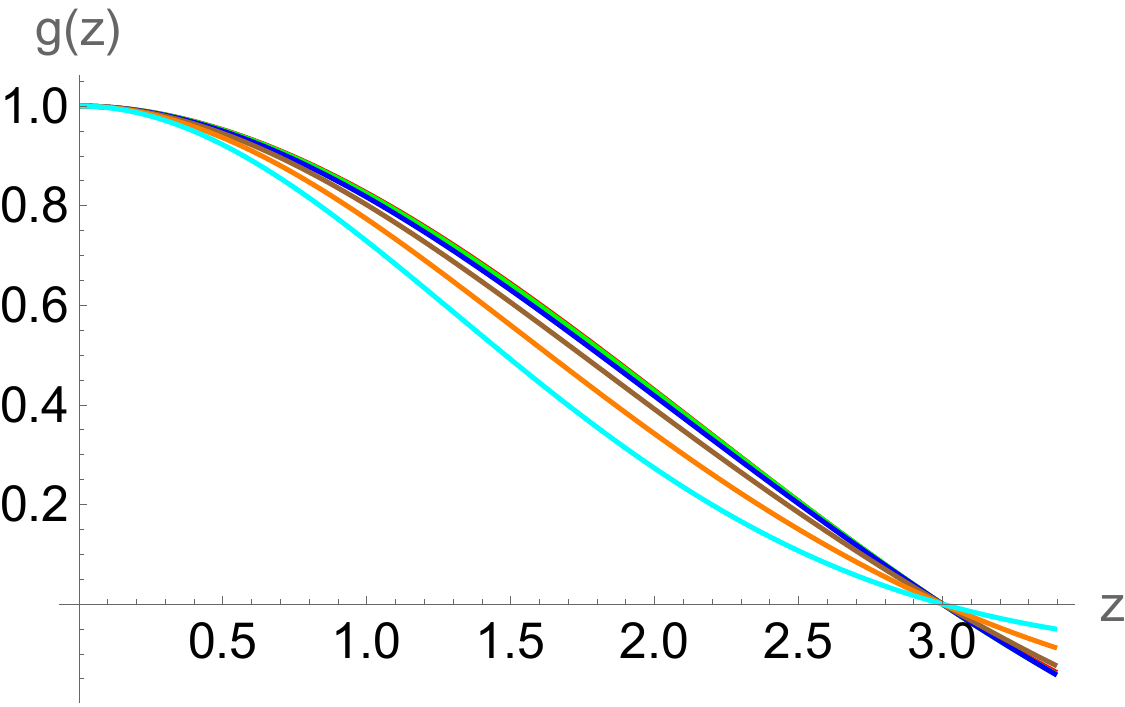}
}
\subfigure[]{
\includegraphics[scale=0.4]{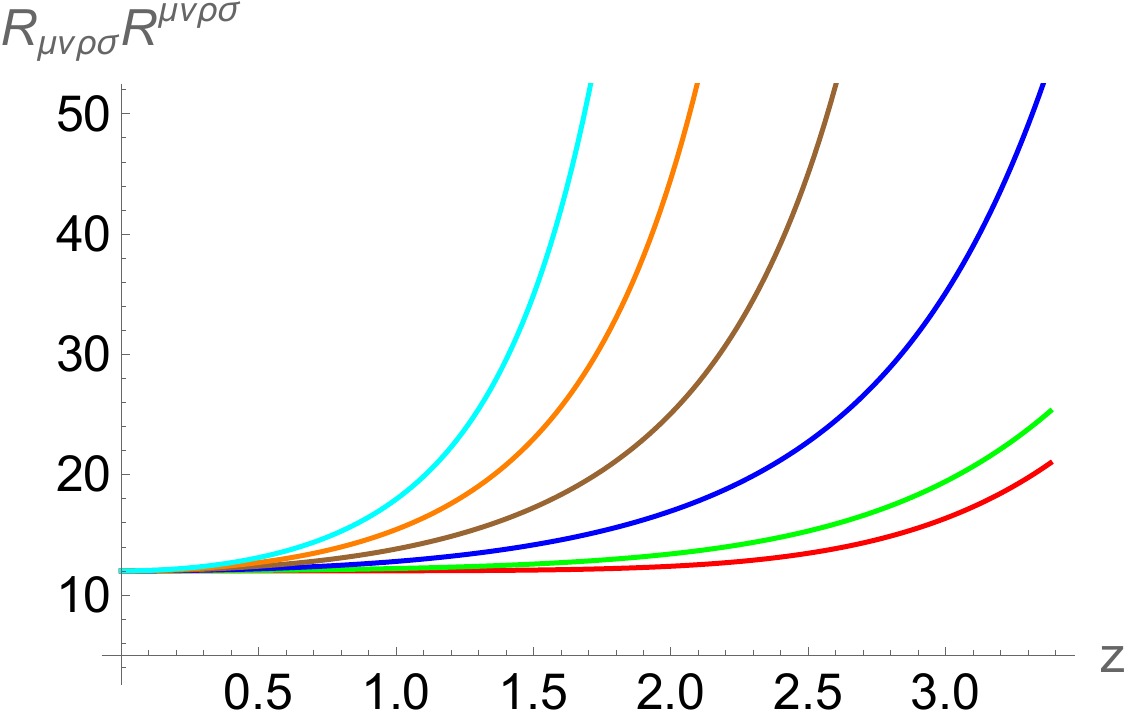}
\label{zvsKretschmannvsazh3qPt2f1}
}
\subfigure[]{
\includegraphics[scale=0.4]{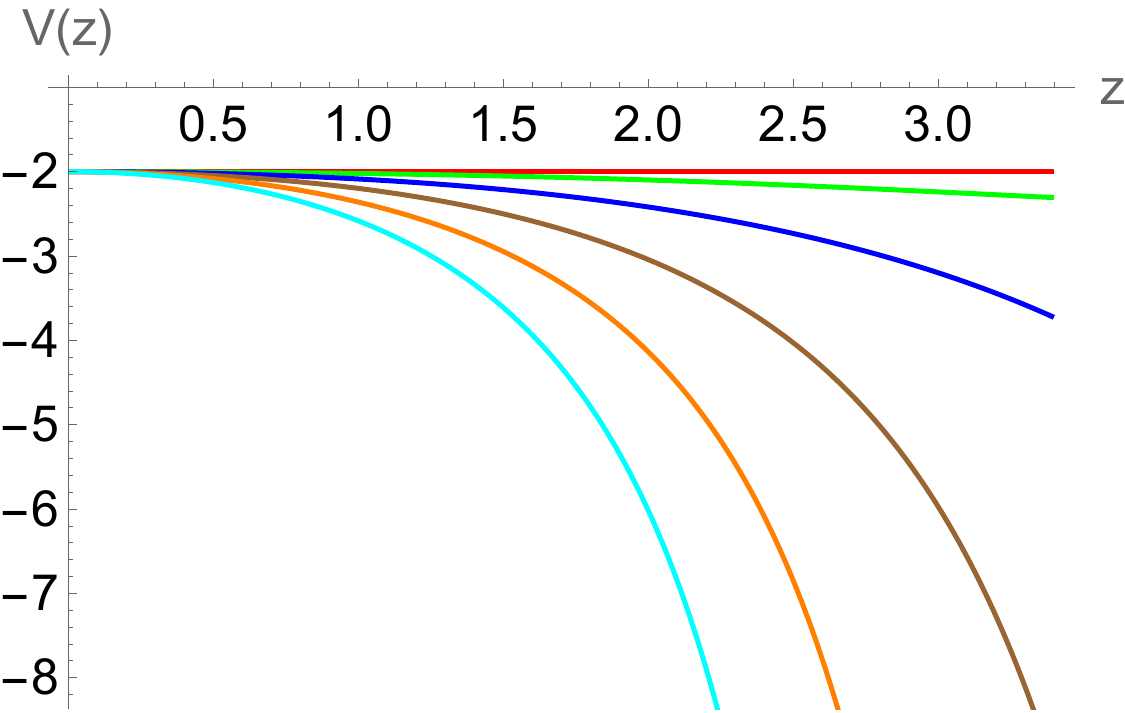}
}
\subfigure[]{
\includegraphics[scale=0.4]{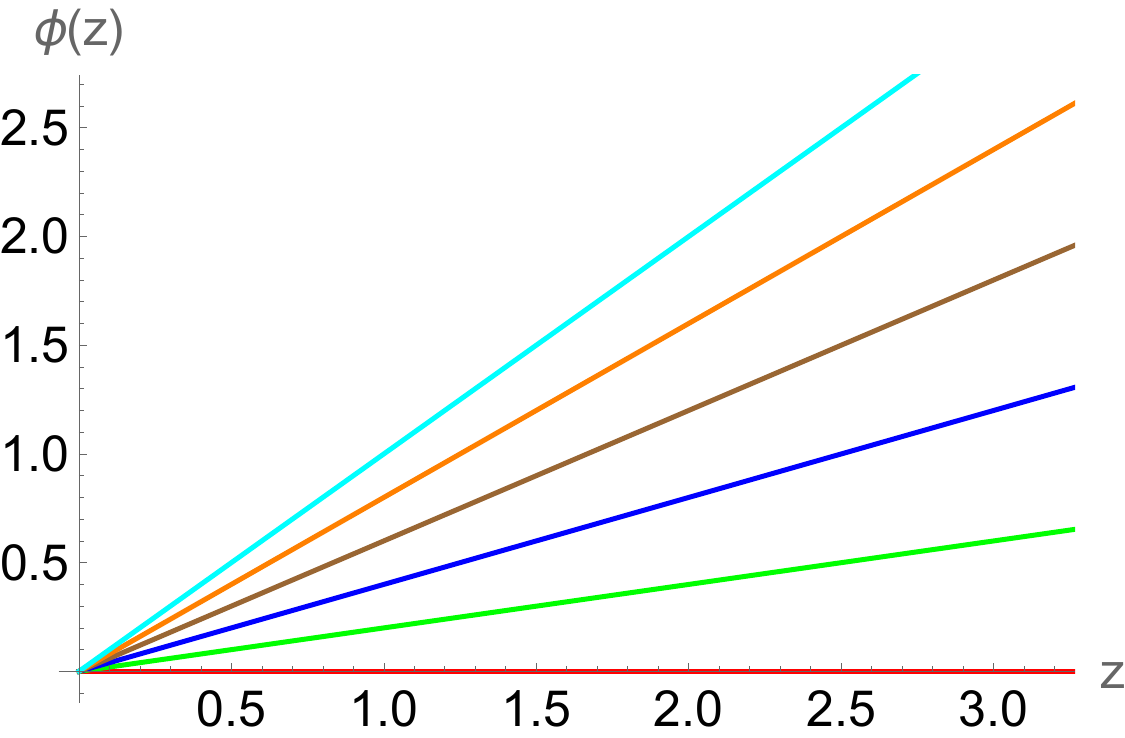}
}
    \caption{The nature of $g(z)$, $R_{\mu\nu\rho\sigma}R^{\mu\nu\rho\sigma}$, $\phi(z)$, and $V(z)$ for different values of hairy parameter $a$. Here $z_{h}=3.0$ and $q_e=0.15$ are used. Red, green, blue, brown, orange, and cyan curves correspond to $a = 0$, $0.1$, $0.2$, $0.3$, $0.4$, and $0.5$, respectively.}
    \label{gvszcase1}
\end{figure}

In Fig.~\ref{gvszcase1}, the radial profile of $g(z)$, Kretschmann scalar $R_{\mu\nu\rho\sigma}R^{\mu\nu\rho\sigma}$, scalar field, and the potential is shown for different values of scalar hair parameter $a$. Note that, at $z=z_{h}$, $g(z)$ changes its sign, indicating the presence of a horizon. This is true for all values of $a$. Similarly, the Ricci and Kretschmann scalars are finite everywhere outside the horizon. The curvature singularity appears only at $z=1/r=\infty$, which is shielded by the horizon. Therefore, there is no additional singularity in the hairy black hole case than those already present at the nonhairy charged BTZ black hole. Note that in the usual three-dimensional Einstein-Maxwell gravity system, the curvature is constant throughout the spacetime for the uncharged BTZ case. The curvature singularity arises only when the charge is added. The same is true for the Einstein-power Maxwell gravity system. However, in the presence of scalar hair, the curvature singularity can arise even when the charge is zero, i.e., the strength of the singularity increases with the scalar hair. In particular, $R_{\mu\nu\rho\sigma}R^{\mu\nu\rho\sigma} \propto z^2$ for the charged BTZ case, whereas $R_{\mu\nu\rho\sigma}R^{\mu\nu\rho\sigma} \propto z^6$ for the hairy case of fractional power law electrodynamics employed here. This can be clearly observed in Fig.~\ref{zvsKretschmannvsazh3qPt2f1}. Interestingly, compared to the Einstein-Maxwell-scalar gravity system, where there is an additional logarithmic singularity at the centre of the black hole ($R_{\mu\nu\rho\sigma}R^{\mu\nu\rho\sigma} \propto z^6\log{z}$) \cite{Priyadarshinee:2023cmi}, the strength of the singularity is milder in the case of Einstein-power Maxwell-scalar gravity system where no such logarithmic singularity arises. This is expected considering that the gauge field gives $\log{z}$ contribution to $g(z)$, thereby giving additional $\log{z}$ contributions to the curvature scalars in the Einstein-Maxwell theory, whereas no such contribution arises in the  Einstein-power Maxwell theory.

Also, the scalar field is finite and real everywhere at and outside the horizon and only goes to zero at the asymptotic boundary. This suggests the presence of a well-behaved hairy black hole solution with Maxwell-like electrodynamics in three dimensions. In the outer horizon area, the potential is similarly regular and limited. The potential asymptotes to $V(z=0)=-2/L^2$ at the boundary for all $a$ and $q_e$. Moreover, provided that the charge $q_e$ is not too large, the potential is also bounded from above by its UV boundary value, i.e., $V(0)\geq V(z)$, hence satisfying the Gubser criterion to have a well-defined boundary field theory \cite{Gubser:2000nd}. However, the said criterion can be violated for higher values of $q_e \gtrsim 5$. In the rest of the work, we will concentrate on only those parameter values for which the Gubser criterion is respected.

Now, let us discuss the thermodynamics of the black hole. For $f(\phi)=1$, the expression of the black hole temperature is given by:
\begin{equation}
T= \frac{a^2 z_h
   e^{a^2 z_h^2}}{2 \pi  \left(e^{a^2 z_h^2}-1\right)} + \frac{q_{e}^{3/2} z_h e^{a^2 z_h^2} \left(3 \sqrt{\pi }
   \text{erf}\left(\sqrt{2} a z_h\right)-3 \sqrt{2 \pi } \text{erf}\left(a
   z_h\right)\right)}{32 \pi  a \left(e^{a^2 z_h^2}-1\right)}\,.
\label{tempcase1}
\end{equation}
The aforementioned expression smoothly reduces to the typical charged BTZ-like expression in the limit when $a\to0$, i.e.,
\begin{equation}
T\big\rvert_{a \rightarrow0} =  \frac{1}{2 \pi  z_h}-\frac{q_e^{3/2} z_h^2}{8 \sqrt{2} \pi }\,,
\label{tempcase1a0}
\end{equation}
which also indicates that the black hole can become extremal when the charge is added to the system, as opposed to the uncharged case.

\begin{figure}[h!]
\begin{minipage}[b]{0.5\linewidth}
\centering
\includegraphics[width=2.8in,height=2.3in]{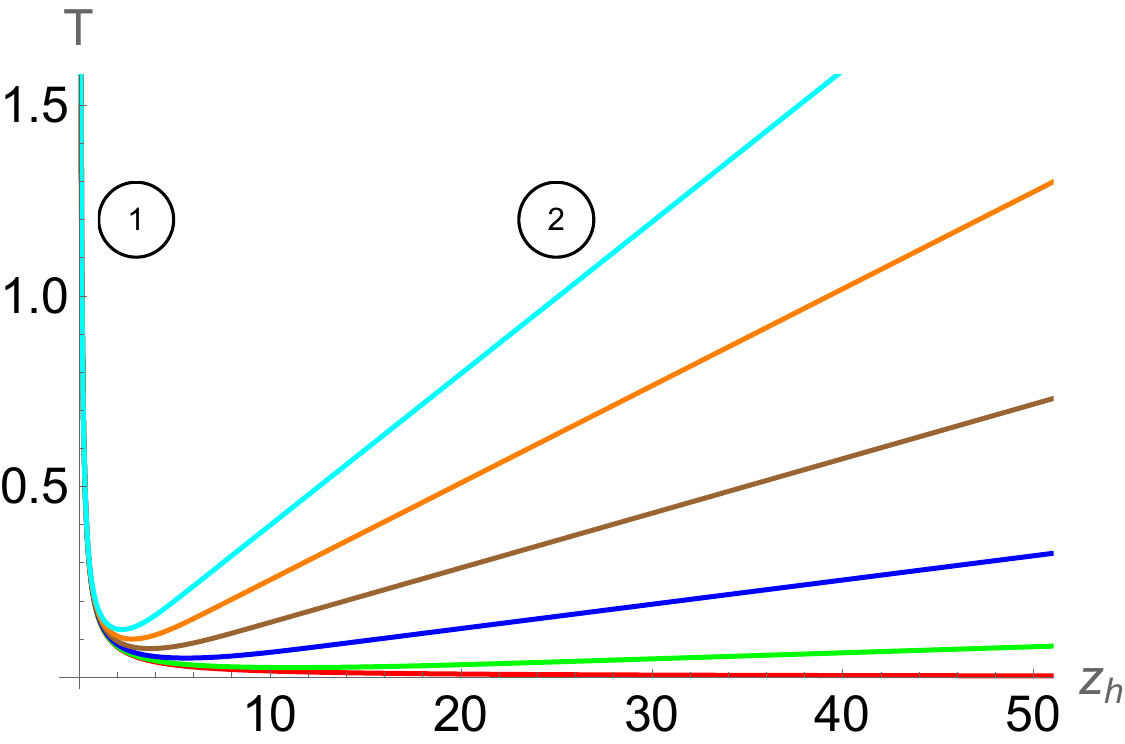}
\caption{ \small Hawking temperature $T$ as a function of horizon radius $z_h$ for various values of $a$. Here $\mu_e=0$ is used. Red, green, blue, brown, orange, and cyan curves correspond to $a=0$, $0.1$, $0.2$, $0.3$, $0.4$, and $0.5$, respectively.}
\label{zhvsTvsamu0f1}
\end{minipage}
\hspace{0.4cm}
\begin{minipage}[b]{0.5\linewidth}
\centering
\includegraphics[width=2.8in,height=2.3in]{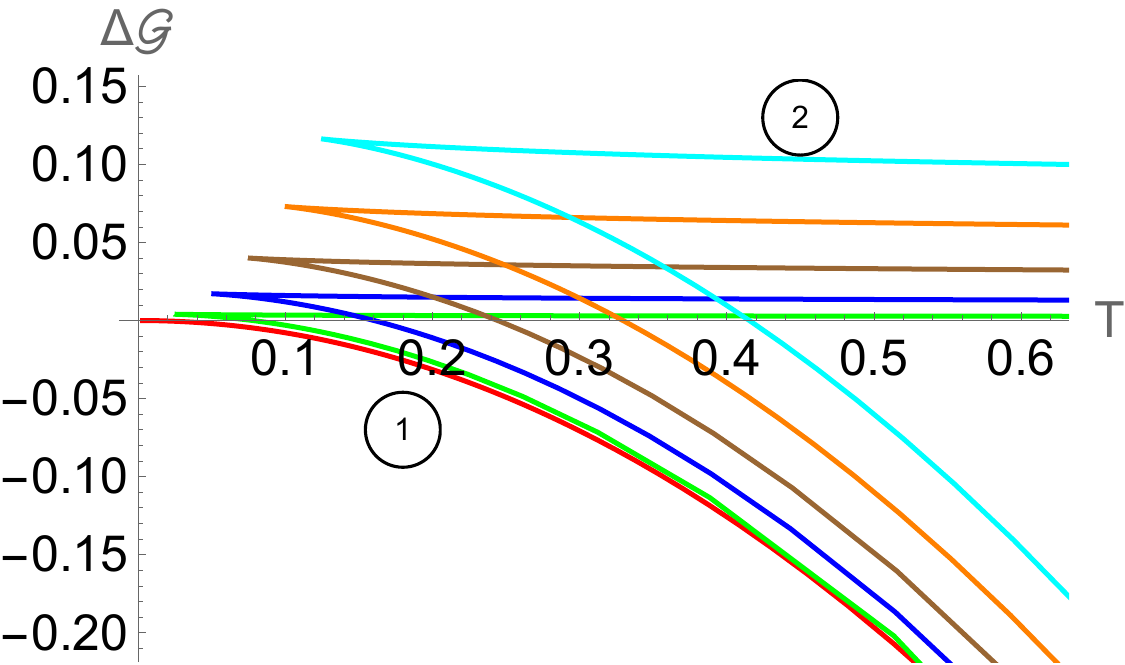}
\caption{\small The Gibbs free energy difference $\Delta\mathcal{G}$ as a function of $T$ for various values of $a$. Here $\mu_e=0$ is used. Red, green, blue, brown, orange, and cyan curves correspond to $a=0$, $0.1$, $0.2$, $0.3$, $0.4$, and $0.5$, respectively.}
\label{TvsGvsamu0f1}
\end{minipage}
\end{figure}

Let us first discuss the black hole thermodynamics in the
grand-canonical ensemble. Fig.~\ref{zhvsTvsamu0f1} illustrates the temperature variation with regard to the (inverse) horizon radius $z_{h}$ for various values of hairy parameter $a$. Here we have kept $\mu_e=0$ fixed, which is also equivalent to $q_e=0$. Observe that for $a=0$ (red line), there is only one black hole phase. The temperature of this phase decreases with $z_h$ and has a positive specific heat. Correspondingly, this black hole phase is thermodynamically stable. The local thermodynamic stability of the hairy black holes will be discussed shortly. This is an expected result considering that for $a=0$ and $\mu_e=0$, our hairy solution reduces to the stable uncharged BTZ black hole, which is thermodynamically stable at all temperatures.

The thermodynamic structure changes drastically when the hairy parameter $a$ is switched on. With finite $a$, there are now two black hole phases at a fixed temperature: a small phase (unstable) and a large phase (stable). These stable and unstable phases are marked by \textcircled{1} and \textcircled{2}, respectively, in Fig.~\ref{zhvsTvsamu0f1}. While the temperature increases with $z_{h}$  for the small-unstable black hole phase, it falls with $z_{h}$  for the large-stable black hole phase (cyan line). The appearance of the small-unstable phase can also be analytically noticed from Eq.~(\ref{tempcase1}). Observe that for $\mu_e=0$, the second term vanishes, and only the first term contributes to the temperature, and this first term increases with $z_h$ for large $z_h$.  Interestingly, unlike the uncharged BTZ black hole, the uncharged hairy black hole phases exist only above a certain minimum temperature $T_{min}$, i.e.,  below $T_{min}$, the hairy black hole phases cease to exist and leaving the thermal-AdS solution as the only remaining feasible phase.  This is true for all finite values of $a$. Importantly, as it generally happens, the occurrence of multivaluedness of the temperature also indicates a possible phase transition in the hairy context.

To further investigate the global thermodynamic stabilities of the above-discussed hairy black hole phases, we need to study their free energy behaviour. The Gibbs free energy $\mathcal{G}$ at a fixed  potential in differential form is related to the black hole entropy as
\begin{eqnarray}
& & d\mathcal{G}=-S_{BH} \ d T\,,
\label{dGSdT}
\end{eqnarray}
which can be used to compute the free energy difference between the black hole and thermal-AdS phases,
\begin{eqnarray}
& & \Delta\mathcal{G} =-\int S_{BH}d T = - \int^{z_h}_{z_\Lambda=\infty} S_{BH} \frac{d T}{d z_h} dz_h \,.
\label{Gibbsfreeenergy}
\end{eqnarray}

In Fig.~\ref{TvsGvsamu0f1}, the Gibbs free energy of the hairy black hole phases is shown.\footnote{Here, we have taken the upper limit of integration $z_\Lambda=10^6$ in the numerical computation.} The colour pattern used here is identical to Fig.~\ref{zhvsTvsamu0f1}. We observe that for all finite values of $a$, there appears a transition temperature $T_{HP}$ at which the Gibbs free energy changes its sign. It suggests the occurrence of a well-known Hawking/Page type phase transition between an uncharged large-stable hairy black hole phase and the thermal-AdS phase at $T_{HP}$. Accordingly, below $T_{HP}$ thermal-AdS is thermodynamically favoured, whereas above $T_{HP}$ large hairy black hole is thermodynamically favoured. Also, the free energy of the small-unstable black hole phase is always higher than the large-stable black hole phase, indicating that the small-unstable black hole phase is always thermodynamically disfavored for the large-stable black hole phase.

\begin{figure}[h!]
\begin{minipage}[b]{0.5\linewidth}
\centering
\includegraphics[width=2.8in,height=2.3in]{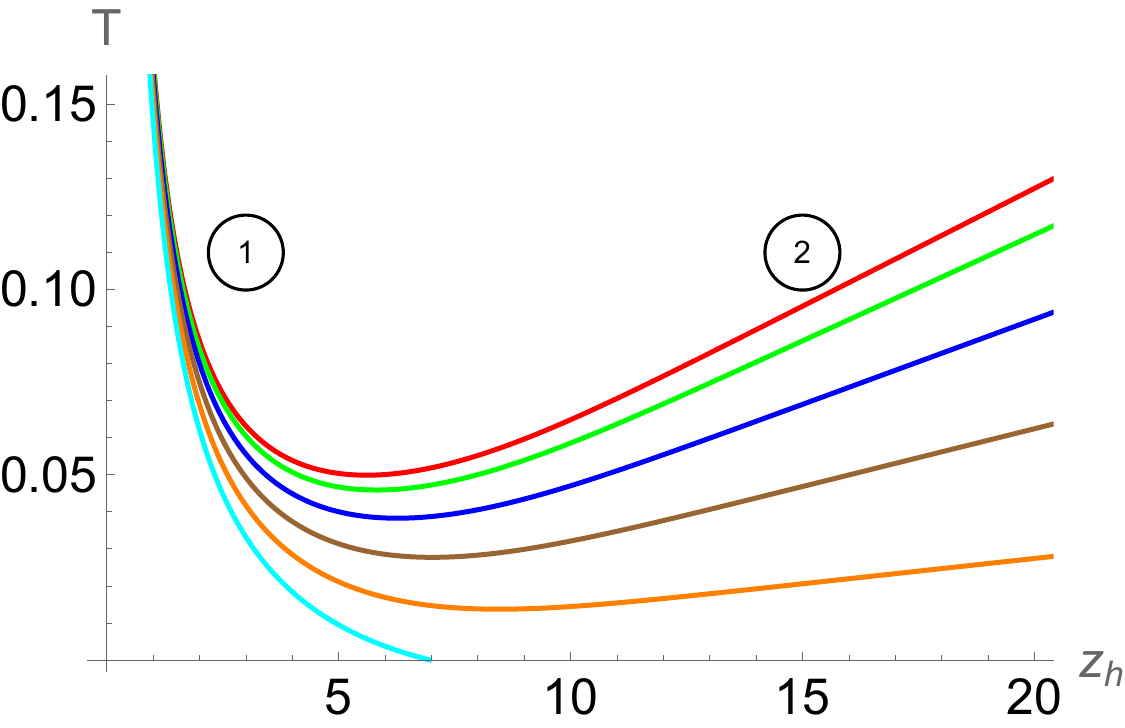}
\caption{ \small Hawking temperature $T$ as a function of horizon radius $z_h$ for various values of chemical potential $\mu_e$.  Here, $a=0.2$ is used. Red, green, blue, brown, orange, and cyan curves correspond to $\mu_e=0$, $0.2$, $0.4$, $0.6$, $0.8$, and $1.0$, respectively.}
\label{zhvsTvsmuaPt2f1}
\end{minipage}
\hspace{0.4cm}
\begin{minipage}[b]{0.5\linewidth}
\centering
\includegraphics[width=2.8in,height=2.3in]{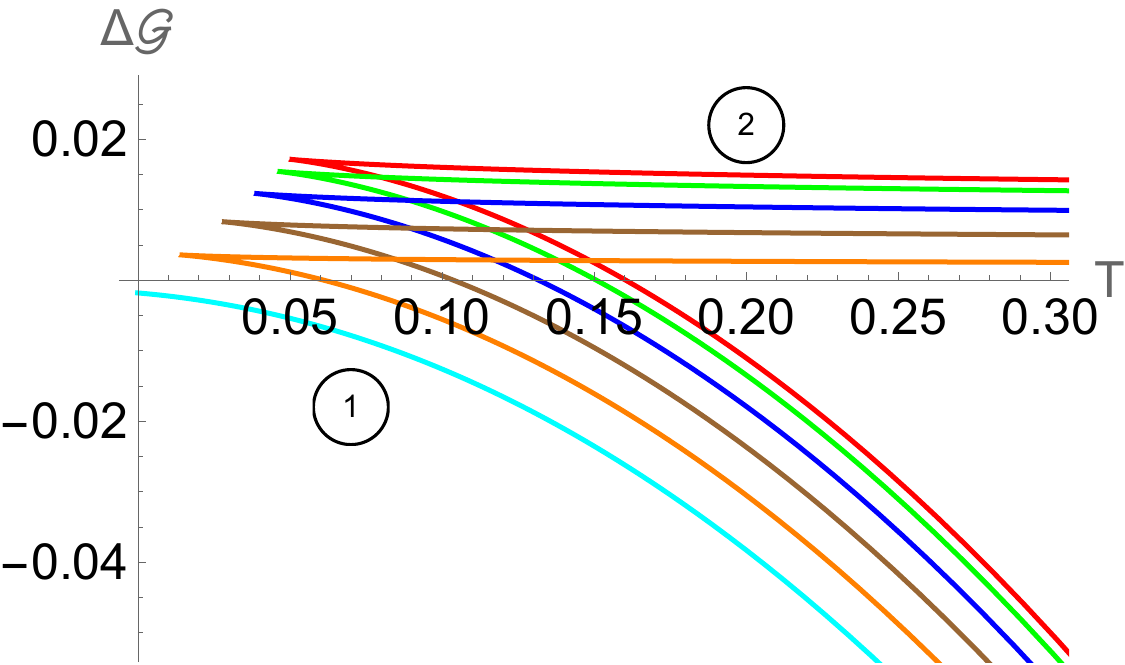}
\caption{\small The Gibbs free energy difference $\Delta\mathcal{G}$ as a function of $T$ for various values of chemical potential $\mu_e$. Here $a=0.2$ is used. Red, green, blue, brown, orange, and cyan curves correspond to $\mu_e=0$, $0.2$, $0.4$, $0.6$, $0.8$, and $1.0$, respectively.}
\label{TvsGvsmuaPt2f1}
\end{minipage}
\end{figure}

The above thermodynamic structure of the hairy black hole gets more interesting as the chemical potential is turned on. Depending upon the relative magnitudes of $\mu_e$ and $a$, not only the hairy black hole can become extremal, but it can also exist in one or two phases. This is illustrated in Fig.~\ref{zhvsTvsmuaPt2f1}. Here, we have presented the results for a particular value of $a = 0.2$; however, analogous results appear for other values of $a$ as well. For small $\mu_e$, like in the $\mu_e=0$ case, there again appear two black hole phases above $T_{min}$, with the large black hole phase (indicated by \textcircled{1}) being thermodynamically more favoured and stable compared to the small black hole phase (indicated by \textcircled{2}) at all temperatures $T>T_{min}$. Therefore, for small $\mu_e$, there again occurs a Hawking/Page phase transition between the large hairy black hole and thermal-AdS phases. This is illustrated in Fig.~\ref{TvsGvsmuaPt2f1}, where one can clearly observe a sign change in the Gibbs free energy as the temperature is varied. However, for large $\mu_e$, this phase transition ceases to exit. For large $\mu_e$, there exists only one stable black hole phase which becomes extremal at some horizon radius $z_h^{ext}$ (cyan line), and the free energy of this black hole phase is always smaller than the thermal-AdS. This result is completely analogous to the charged BTZ black hole. For the non-hairy charged BTZ black hole (with Coulomb-like potential), the extremal horizon radius can be found from Eq.~(\ref{tempcase1a0}). It occurs at $z_h^{ext}=4(2)^{1/6}/\mu_e$. Whereas for the hairy black hole case, the magnitude of this $z_h^{ext}$ increases with $a$. These results further imply that irrespective of the temperature, at least one black hole phase always exists and remains stable for the charged case when $\mu_e$ is relatively large. Our whole analysis, therefore, suggests that for a fixed value of $a$, there exists a critical chemical potential $\mu_e^{c}$ below which the Hawking/Page phase transition between the thermal-AdS and hairy black hole phases takes place, whereas no such phase transition appears above $\mu_e^{c}$.

\begin{figure}[h!]
\begin{minipage}[b]{0.5\linewidth}
\centering
\includegraphics[width=2.8in,height=2.3in]{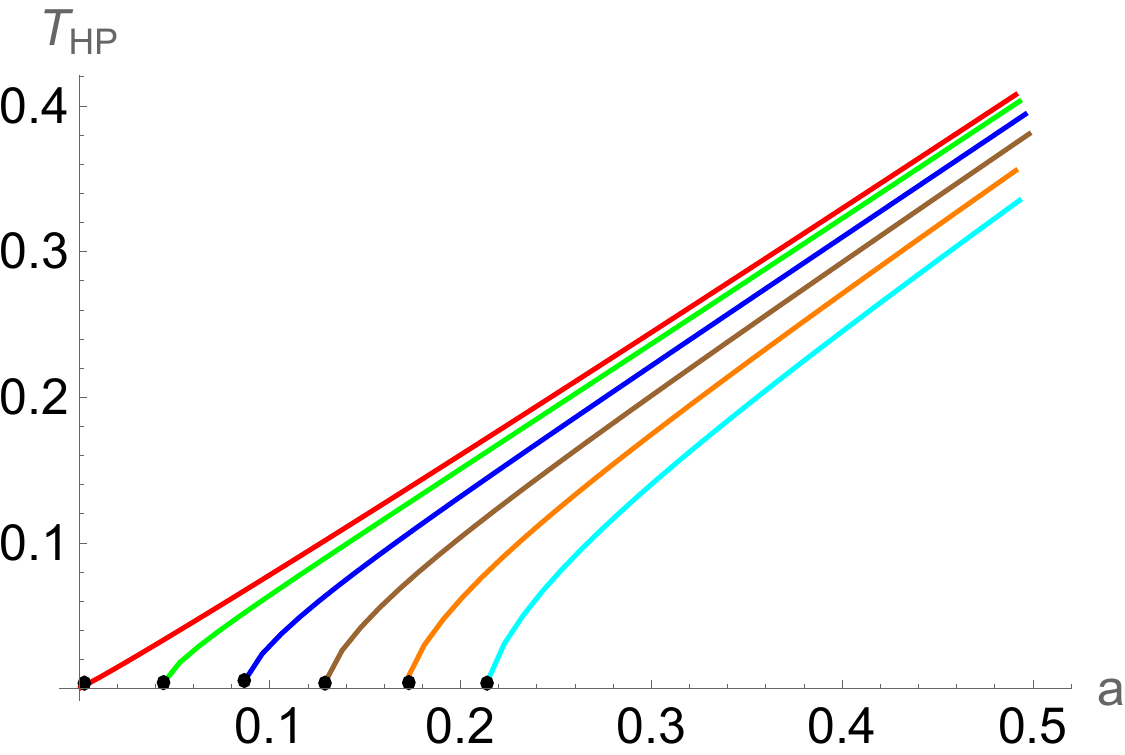}
\caption{ \small Hawking/Page phase transition temperature $T_{HP}$ as a function of $a$ for various values of chemical potential $\mu_{e}$. Red, green, blue, brown, orange, and cyan curves correspond to $\mu_{e}=0$, $0.2$, $0.4$, $0.6$, $0.8$, and $1$, respectively. The black dots indicate the critical hairy parameter $a_c$. }
\label{CritTvsaf1fixedmu}
\end{minipage}
\hspace{0.4cm}
\begin{minipage}[b]{0.5\linewidth}
\centering
\includegraphics[width=2.8in,height=2.3in]{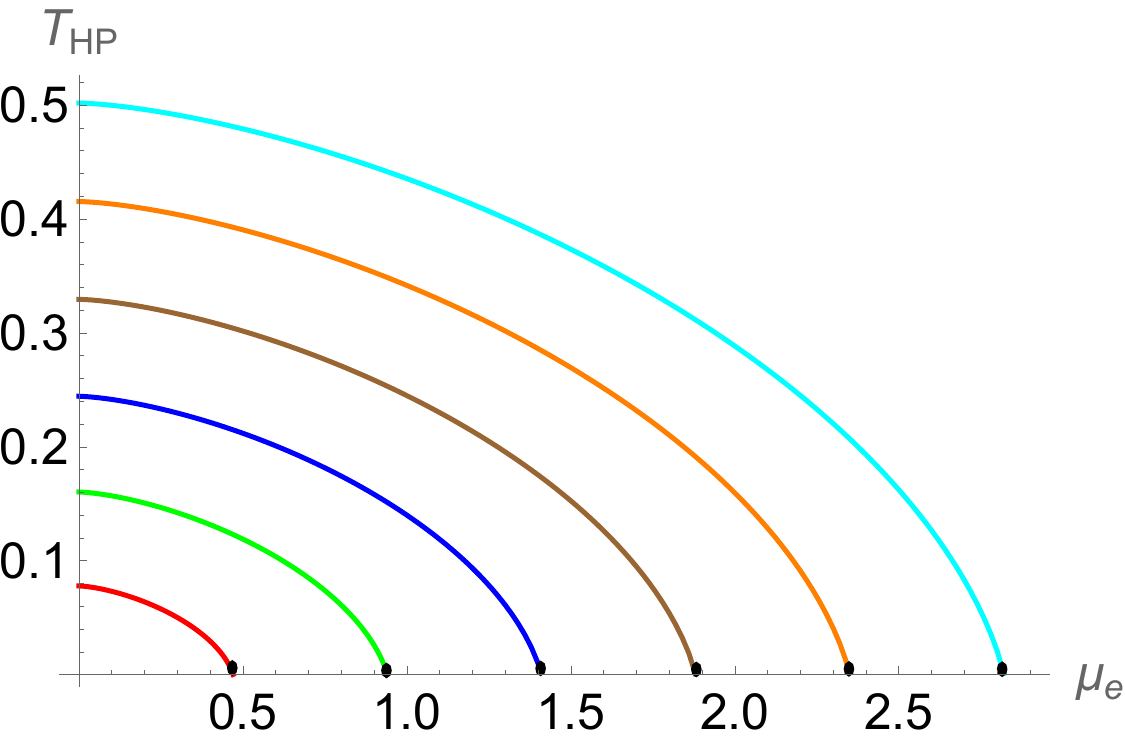}
\caption{\small Hawking/Page phase transition temperature $T_{HP}$ as a function of chemical potential $\mu_{e}$ for various values of $a$. Red, green, blue, brown, orange, and cyan curves correspond to $a=0.1$, $0.2$, $0.3$, $0.4$, $0.5$, and $0.6$, respectively.  The black dots indicate the critical chemical potential $\mu_{e}^{c}$.}
\label{CritTvsmuf1fixeda}
\end{minipage}
\end{figure}

We further investigate the dependence of $T_{HP}$ on $a$ and $\mu_e$. Figs. \ref{CritTvsaf1fixedmu} and \ref{CritTvsmuf1fixeda} depict the overall dependence of $T_{HP}$ on these variables. Our finding shows that when $a$ increases, $T_{HP}$ shows a monotonically rising tendency. Specifically, the transition temperature rises with $a$ while falling with $\mu_e$. Although $T_{HP}$ rises with $a$ for every $\mu_e$, it should be noted that, unlike the $\mu_e = 0$ case, the slope of the $a$ vs $T_{HP}$ line is not constant for finite $\mu_e$. Our results, therefore, suggest that the possibility of Hawking/Page phase transition in the hairy context increases for large $a$ and small $\mu_e$ values. It also implies that for large chemical potential, one needs a higher value of $a$ to observe the Hawking/Page phase transition. This, in turn, implies the existence of critical hairy parameter $a_c$ below which no Hawking/Page phase transition takes place in the fixed $\mu_e$ grand-canonical ensemble.

\begin{figure}[h!]
\begin{minipage}[b]{0.5\linewidth}
\centering
\includegraphics[width=2.8in,height=2.3in]{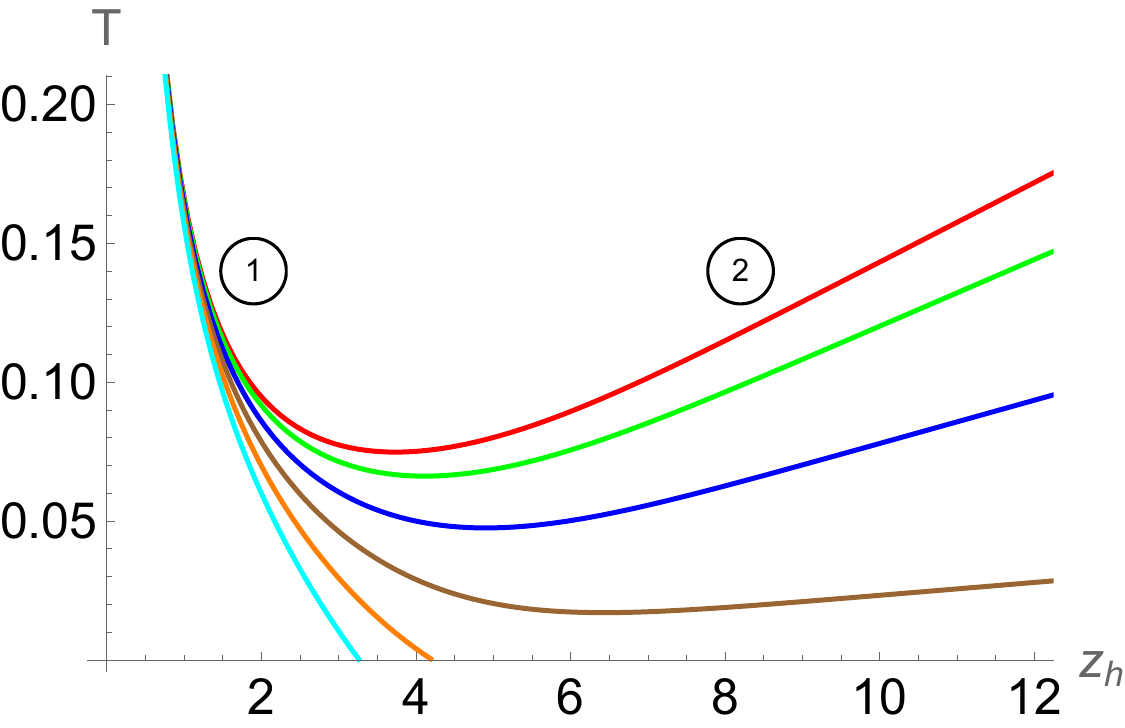}
\caption{ \small Hawking temperature $T$ as a function of horizon radius $z_h$ for various values of charge $q_e$.  Here, $a=0.3$ is used. Red, green, blue, brown, orange, and cyan curves correspond to $q_e=0$, $0.1$, $0.2$, $0.3$, $0.4$, and $0.5$, respectively.}
\label{zhvsTvsqeaPt3f1}
\end{minipage}
\hspace{0.4cm}
\begin{minipage}[b]{0.5\linewidth}
\centering
\includegraphics[width=2.8in,height=2.3in]{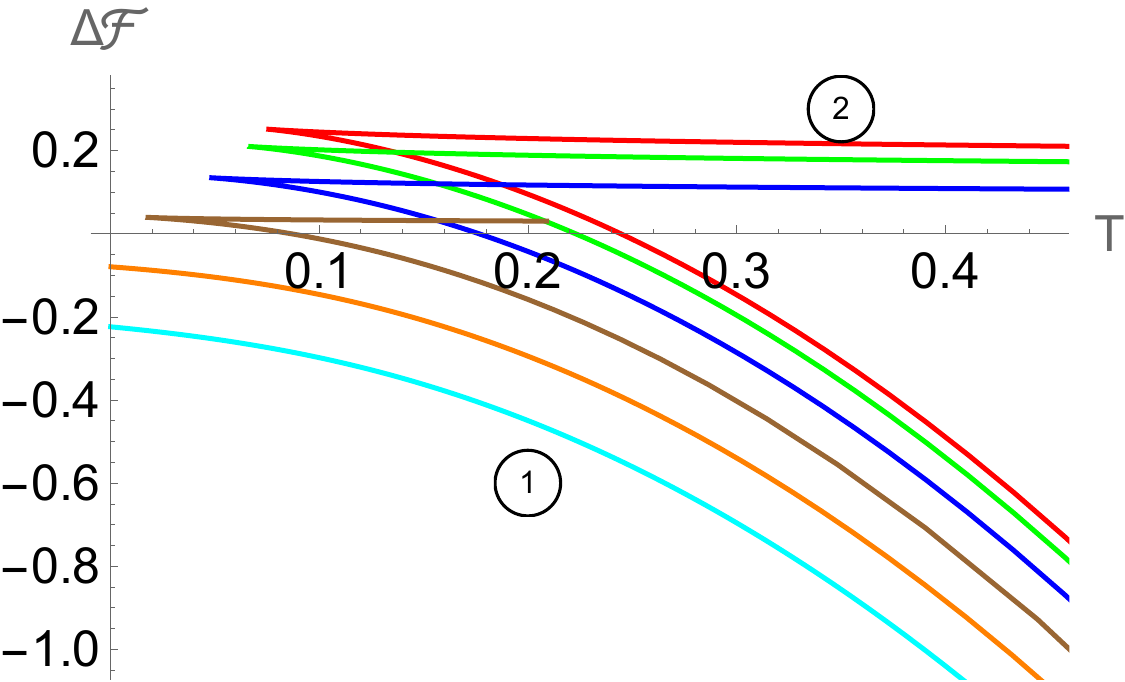}
\caption{\small The Helmholtz free energy difference $\Delta\mathcal{F}$ as a function of $T$ for various values of charge $q_e$. Here $a=0.3$ is used. Red, green, blue, brown, orange, and cyan curves correspond to $q_e=0$, $0.1$, $0.2$, $0.3$, $0.4$, and $0.5$, respectively.}
\label{TvsFvsqeaPt3f1}
\end{minipage}
\end{figure}

Having discussed the thermodynamic structure of the hairy black hole in the grand-canonical ensemble, we now move on to discuss it in the canonical ensemble. We find that the thermodynamic results in the canonical ensemble are quite similar to the grand-canonical ensemble. The case $q_e=0=\mu_e$ is already discussed above. The results for finite $q_e$ are shown in Figs.~\ref{zhvsTvsqeaPt3f1} and \ref{TvsFvsqeaPt3f1}. In the canonical ensemble as well, depending upon the magnitude of $a$, there appear two black hole phases for small $q_e$, whereas only one black hole phase appears for large $q_e$. While the temperature increases with $z_{h}$  for the unstable black hole phase, it falls with $z_{h}$  for the stable black hole phase. The temperature expression shows that for $q_e\neq0$ and very small values of $a$, there is only one stable branch of the black hole, and it becomes extremal at some horizon radius $z_{h}^{ext}$ (orange and cyan lines). These findings also suggest that, for the fixed charged case, at least one stable black hole branch always exists when $a$ is relatively small. While keeping $q_e$ fixed, the temperature starts to rise with $z_{h}$ for large $a$ values. Notably, the hairy black hole phases are restricted to temperatures over a certain threshold $T_{min}$, in contrast to the BTZ black hole. This suggests the possibility of Hawking/Page transition between large hairy black hole phase and thermal-AdS as $a$ increases in the canonical ensemble as well. This is indeed the case, as can be explicitly observed from the free energy behavior shown in Fig.~\ref{TvsFvsqeaPt3f1}. The Helmholtz free energy difference between the black hole and thermal-AdS phases can be computed from the analogous differential first law
\begin{eqnarray}
& & d\mathcal{F}=-S_{BH} \ d T\,, \nonumber\\
& & \Delta\mathcal{F} =\int_{z_h}^{z_\Lambda=\infty} S_{BH} \frac{d T}{d z_h} dz_h \,.
\label{Helmholtzfreeenergy}
\end{eqnarray}

\begin{figure}[h!]
\begin{minipage}[b]{0.5\linewidth}
\centering
\includegraphics[width=2.8in,height=2.3in]{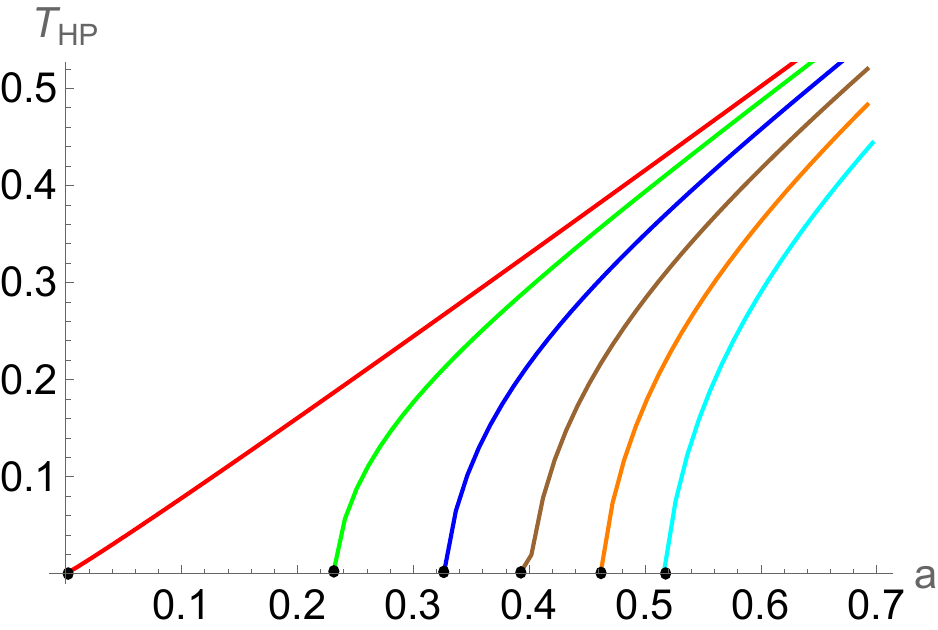}
\caption{ \small Hawking/Page phase transition temperature $T_{HP}$ as a function of $a$ for various values of $q_e$. Red, green, blue, brown, orange, and cyan curves correspond to $q_e=0$, $0.2$, $0.4$, $0.6$, $0.8$, and $1.0$, respectively. The black dots indicate the critical hairy parameter $a_c$.}
\label{CritTvsaf1}
\end{minipage}
\hspace{0.4cm}
\begin{minipage}[b]{0.5\linewidth}
\centering
\includegraphics[width=2.8in,height=2.3in]{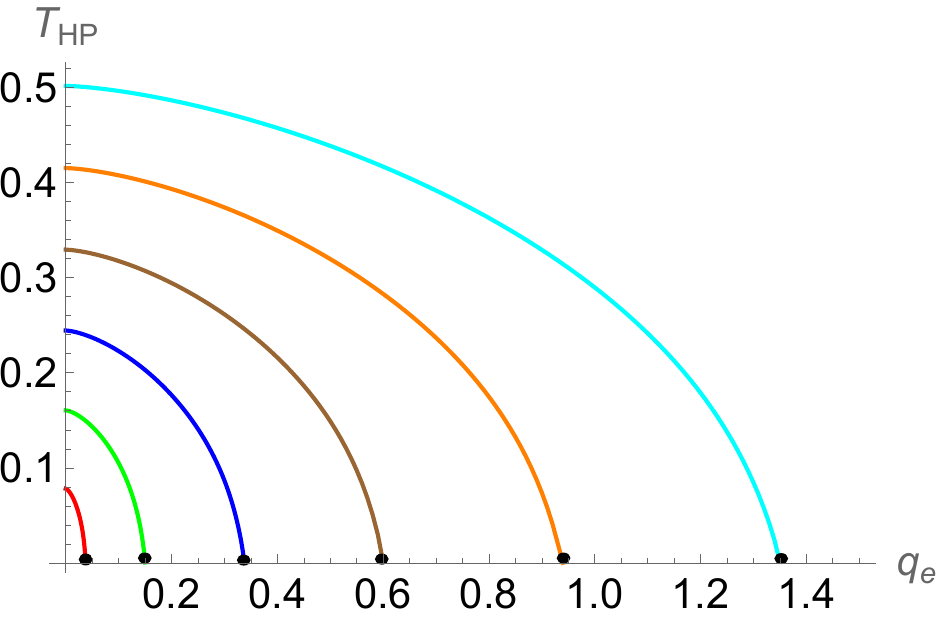}
\caption{\small Hawking/Page phase transition temperature $T_{HP}$ as a function of $q_e$ for various values of $a$. Red, green, blue, brown, orange, and cyan curves correspond to $a=0.1$, $0.2$, $0.3$, $0.4$, $0.5$, and $0.6$, respectively. The black dots indicate the critical charge $q_{e}^{c}$. }
\label{CritTvsqf1}
\end{minipage}
\end{figure}

The transition temperature $T_{HP}$ again depends nontrivially on $a$ and $q_e$. Figs.~\ref{CritTvsaf1} and \ref{CritTvsqf1} show the full illustration of this dependency. Our analysis shows that when $a$ changes, $T_{HP}$ shows a monotonically rising tendency. Specifically, the transition temperature rises with $a$ while falling with $q_e$. Although $T_{HP}$ rises with $a$ for every $q_e$, it should be noted that, unlike the situation where $q_e=0$, the slope of the $a$ vs $T_{HP}$ line is not constant for $q_e\neq0$. Overall, our analysis suggests that there exists a critical value $a_c$ of the hairy parameter in the fixed charge ensemble as well. The charged hairy black hole undergoes the Hawking/Page phase transition above this critical value, while below $a_{c}$, no such phase transition takes place.\footnote{Here we like to emphasize that the Helmholtz free energy difference can not be computed from Eq.~(\ref{Helmholtzfreeenergy}) for nonhairy black holes corresponding to $a\rightarrow 0$. This is because the integrand in Eq.~(\ref{Helmholtzfreeenergy}) contains $z_{h}^{3}$ and $\log{z_h}$ terms which give diverging contributions to the free energy difference in the upper limit of integration. For this reason, here as well as in the subsequent sections, we discuss thermodynamic results only for the hairy case in the canonical ensemble.}

Let us emphasize that the obtained hairy black holes are also locally stable. The local stability corresponds to the response of the equilibrium system under a small fluctuation in thermodynamical variables and is established by the positivity of the specific heat at a constant chemical potential $C_{\mu_e}=T(\partial S_{BH}/\partial T)|_{\mu_e}$ or charge $C_{q_e}=T(\partial S_{BH}/\partial T)|_{q_e}$ in the grand-canonical and canonical ensembles, respectively. Since $S_{BH} \propto z_{h}^{-1}$, it is straightforward to see from Figs.~\ref{zhvsTvsamu0f1} and \ref{zhvsTvsqeaPt3f1} that the slope of the $S_{BH}-T$ plane is always positive in the thermodynamically favoured hairy black hole phase \textcircled{1}. Accordingly, $C_{\mu_e}>0$ and $C_{q_e}>0$ in the favoured hairy black hole phase, indicating the local stability of hairy black holes. Similarly, $C_{\mu_e}$ and $C_{q_e}$ are negative in the thermodynamically disfavored hairy black hole phase \textcircled{2}.

\section{\label{sec:level1} Hairy Black hole solution with Coupling $f(\phi) = e^{\phi}$ }
\begin{figure}[ht]
\subfigure[]{
\includegraphics[scale=0.4]{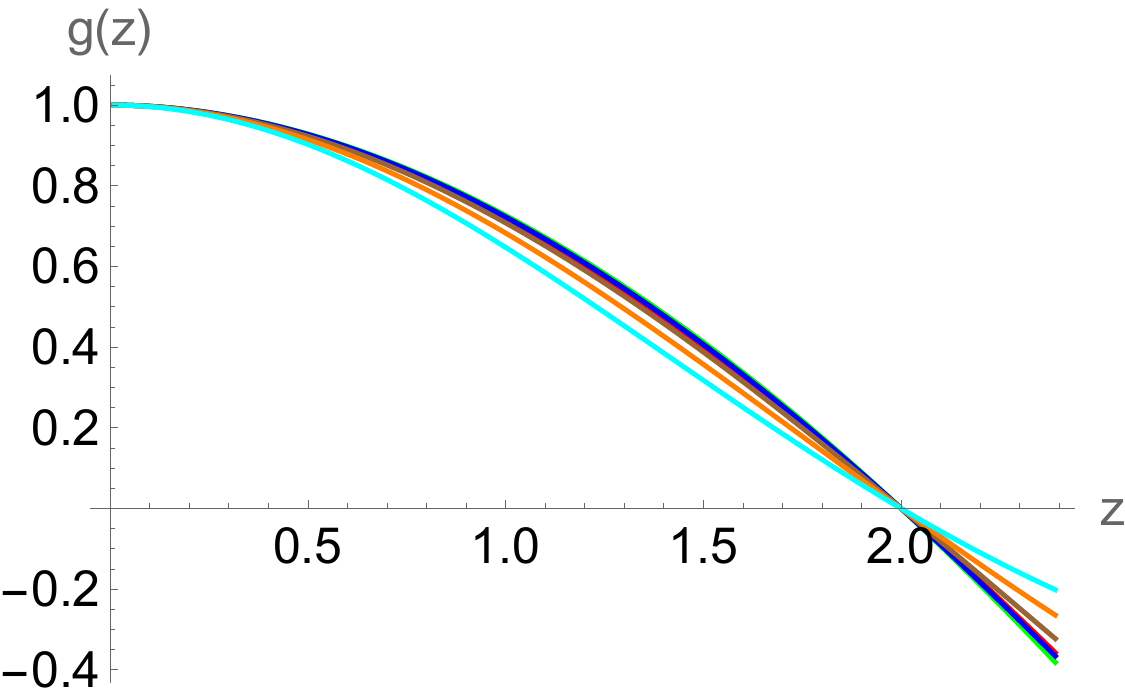}
}
\subfigure[]{
\includegraphics[scale=0.4]{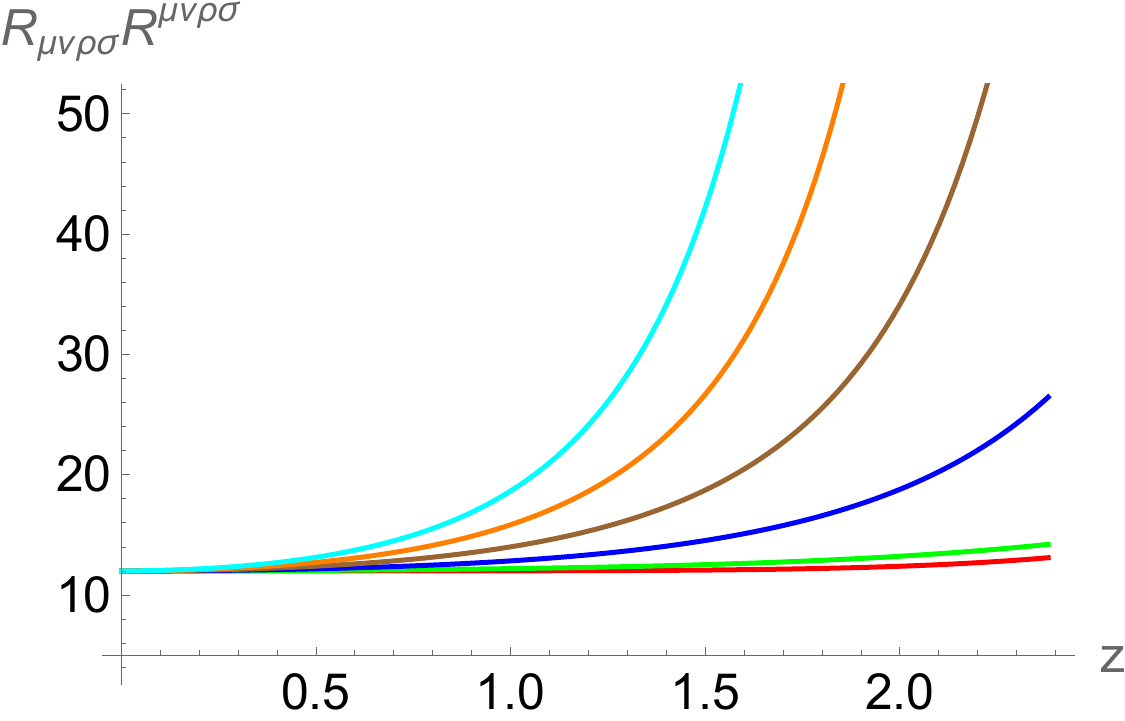}
\label{zvsKretschmannvsazh3qPt2case1}
}
\caption{The nature of $g(z)$ and $R_{\mu\nu\rho\sigma}R^{\mu\nu\rho\sigma}$ for different values of hairy parameter $a$. Here $z_{h}=2.0$ and $q_e=0.2$ are used. Red, green, blue, brown, orange, and cyan curves correspond to $a = 0$, $0.1$, $0.2$, $0.3$, $0.4$, and 0.5, respectively.}
    \label{gvszcase2}
\end{figure}
In this section, we investigate the geometrical and thermodynamical structure of the charged hairy black hole solutions for the exponential coupling function $f(\phi) = e^{\phi}$. The form of $A(z)=-a^2z^2$ is the same as in the previous section. Therefore, the solution of the scalar field will remain the same. This indicates that the scalar field remains regular, finite, and well-behaved everywhere outside the horizon for this coupling as well. From the equation \eqref{electricfield}, we get gauge field solution
\begin{equation}
B_{t}(z)=  \frac{q_e \sqrt{\pi } e^4}{\sqrt{2} a} \left(\text{erf}(2+a z_h)-\text{erf}\left(2+a z\right)\right)\,,
\label{gaugefield2}
\end{equation}
with relation between $\mu_e$ and $q_e$ as
\begin{equation}
q_e= \sqrt{\frac{2}{\pi }} \frac{a \mu_e}{e^4
   \left(\text{erf}\left(2+a z_h\right) - \text{erf}(2)\right)}\,.
\label{gaugefield2}
\end{equation}
Notice that in the limit $a\to 0$, $B_t(z)$ again reduces to the Coulomb-like potential. In a similar manner, we obtain the following expression for $g(z)$:
\begin{eqnarray}
g(z)  = \frac{1-e^{a^2 \left(z_h^2-z^2\right)}}{1-e^{a^2 z_h^2}} +  \frac{3 \sqrt{\pi } q_e^{3/2} \left(\sqrt{2} e^2
   \text{erf}(2)-\text{erf}\left(\sqrt{2}\right)\right) e^{2-a^2 z^2}
    \left(e^{a^2 z_h^2 - e^{a^2 z^2}}\right)}{16 a^3 \left(e^{a^2
   z_h^2}-1\right)} \nonumber\\
  +\frac{3 q_e^{3/2} \sqrt{\pi} e^{4-a^2 z^2} \left(e^{a^2 z^2}-1\right)
    \text{erf}\left(2+a z_h\right)}{8 \sqrt{2} a^3 \left(e^{a^2
   z_h^2}-1\right)} +\frac{3 e^2 \sqrt{\pi } q_e^{3/2} \text{erf}\left(\sqrt{2} (a z+1)\right)}{16a^3}  \nonumber\\
 -\frac{3 q_e^{3/2} \sqrt{\pi } \left(e^{a^2 z^2}-1\right)  e^{2+a^2 z_h^2-a^2
   z^2} \text{erf}\left(\sqrt{2} \left(a z_h+1\right)\right)}{16 a^3
   \left(e^{a^2 z_h^2}-1\right)}  - \frac{3 \sqrt{\pi} e^{4-a^2 z^2} q_e^{3/2} \text{erf}(2+a z)}{8 \sqrt{2} a^3} \,,
\label{g(z)2}
\end{eqnarray}
which again, under the limit $a\to0$, reduces to the standard charged BTZ black hole like expressions with Coulomb-like potential. In Fig.~\ref{gvszcase2}, the behavior of $g(z)$ and the Kretschmann scalar $R_{\mu\nu\rho\sigma}R^{\mu\nu\rho\sigma}$ is illustrated. The spacetime exhibits a horizon at $z_h$ and does not contain any additional singularity, thereby emphasizing the well-behaved nature of the obtained hairy solution. The Ricci scalar is similarly finite and well-behaved everywhere outside the horizon. The hair parameter again non-trivially modifies the Kretschmann and Ricci scalars, implying that the spacetime curvature depends non-trivially on the hairy parameter and is no longer a constant. Similarly, the potential asymptotes to a constant value $V(z)|_{z\rightarrow 0} = 2\Lambda$ at the AdS boundary and is bounded from above.

Let us now talk about the thermodynamics of this black hole. For the coupling function $f(\phi) = e^{\phi}$, the expression of the black hole temperature is
\begin{eqnarray}
T & = & \frac{a^2 z_h
   e^{a^2 z_h^2}}{2 \pi  \left(e^{a^2 z_h^2}-1\right)} + \frac{3 q_{e}^{3/2} z_h e^{a^2 z_h^2+4}
   \left(\text{erf}(2)-\text{erf}\left(a z_h+2\right)\right)}{16 \sqrt{2 \pi }
   a \left(e^{a^2 z_h^2}-1\right)} \nonumber\\
& &  + \frac{3 q_{e}^{3/2} z_h e^{a^2 z_h^2+2} \left(\text{erf}\left(\sqrt{2}
   \left(a z_h+1\right)\right)-\text{erf}\left(\sqrt{2}\right)\right)}{32
   \sqrt{\pi } a \left(e^{a^2 z_h^2}-1\right)} \,,
\label{tempcase2}
\end{eqnarray}
which also reduces to Eq.~(\ref{tempcase1a0}) in the limit $a\rightarrow 0$ and to the uncharged BTZ expression in the limit $a\rightarrow 0$ and $q_e\rightarrow 0$.

Let us again first discuss the black hole thermodynamics in the grand-canonical ensemble. The thermodynamic structure of the hairy black hole with $f(\phi) = e^{\phi}$ coupling remains quite similar to the $f(\phi) = 1$ coupling for the fixed chemical potential ensemble. Notice that for $\mu_e=0$, the Einstein-power Maxwell-scalar gravity action becomes the same for both $f(\phi)=1$ and $f(\phi) = e^{\phi}$ couplings, which in turn produces identical thermodynamic structure for both couplings for $\mu_e=0$. Consequently, for $f(\phi) = e^{\phi}$ coupling as well, there exists a thermodynamically stable hairy black hole phase which undergoes a phase transition to thermal-AdS phase as the temperature is lowered, i.e., the Hawking/page phase transition continues to exist, with the thermal-AdS
phase dominating the structure at lower temperatures whereas a large stable hairy black hole phase dominates the phase
structure at higher temperatures. For $\mu_e=0$, the phase diagram is essentially identical to Fig.~\ref{TvsGvsamu0f1}.

\begin{figure}[h!]
\begin{minipage}[b]{0.5\linewidth}
\centering
\includegraphics[width=2.8in,height=2.3in]{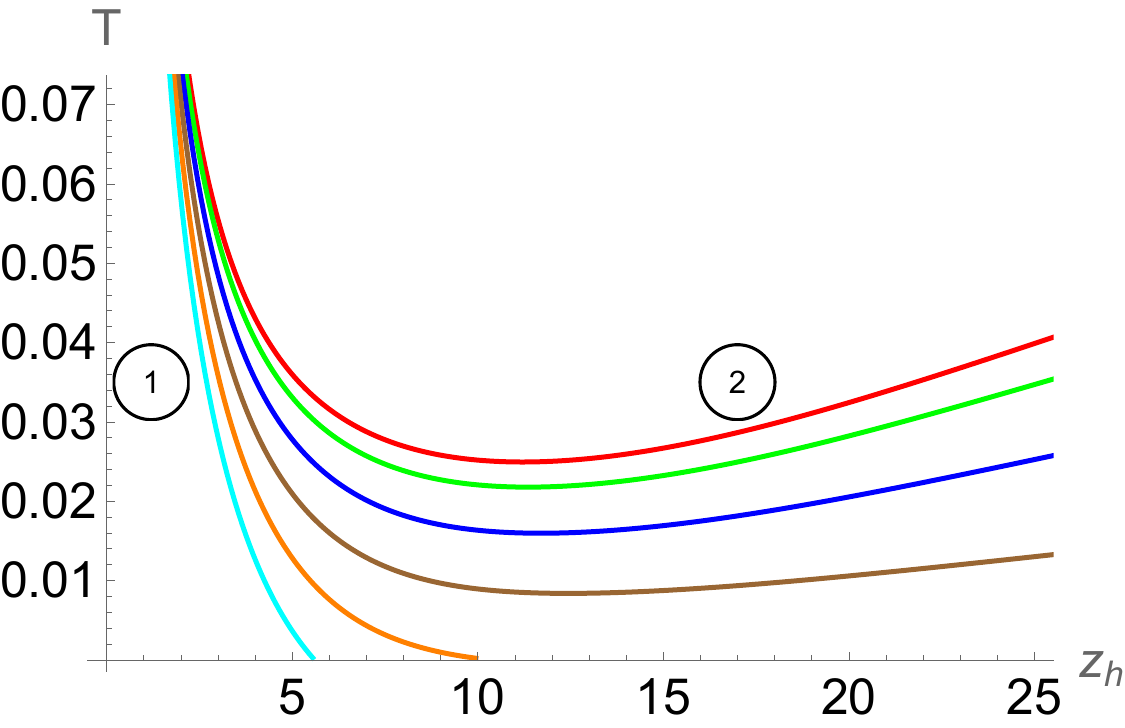}
\caption{ \small Hawking temperature $T$ as a function of horizon radius $z_h$ for various values of chemical potential $\mu_e$.  Here $a=0.1$ is used. Red, green, blue, brown, orange, and cyan curves correspond to $\mu_e=0$, $0.2$, $0.4$, $0.6$, $0.8$, and $1.0$, respectively.}
\label{zhvsTvsmuaPt1f2}
\end{minipage}
\hspace{0.4cm}
\begin{minipage}[b]{0.5\linewidth}
\centering
\includegraphics[width=2.8in,height=2.3in]{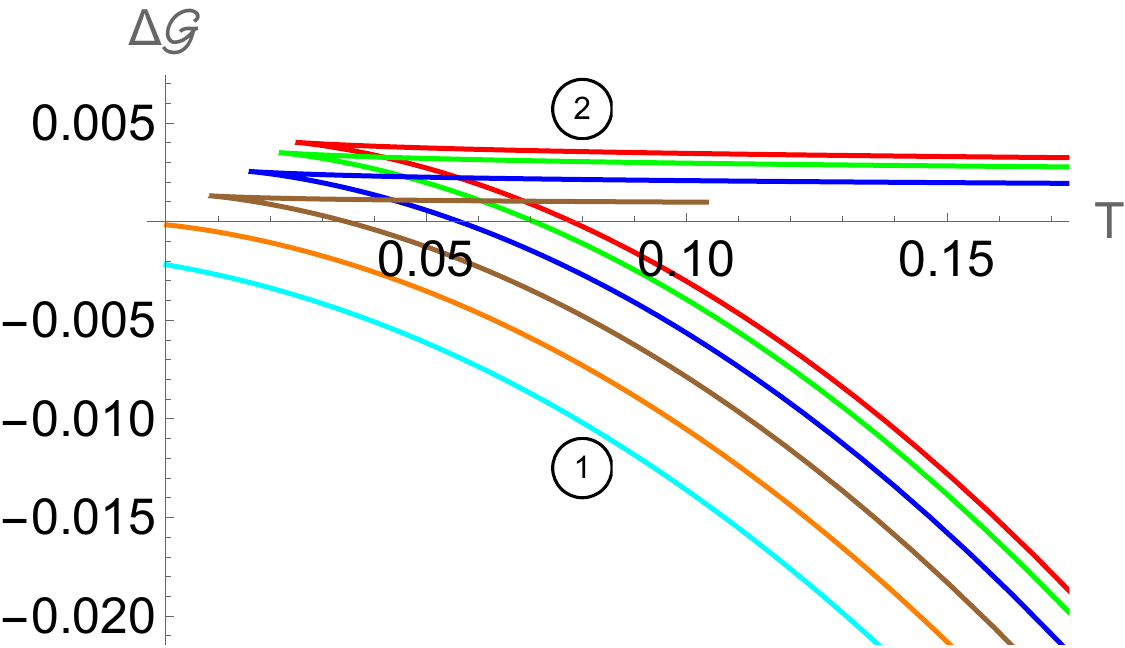}
\caption{\small The Gibbs free energy difference $\Delta\mathcal{G}$ as a function of $T$ for various values of chemical potential $\mu_e$. Here, $a=0.1$ is used. Red, green, blue, brown, orange, and cyan curves correspond to $\mu_e=0$, $0.2$, $0.4$, $0.6$, $0.8$, and $1.0$, respectively.}
\label{TvsGvsmuaPt1f2}
\end{minipage}
\end{figure}
\begin{figure}[h!]
\begin{minipage}[b]{0.5\linewidth}
\centering
\includegraphics[width=2.8in,height=2.3in]{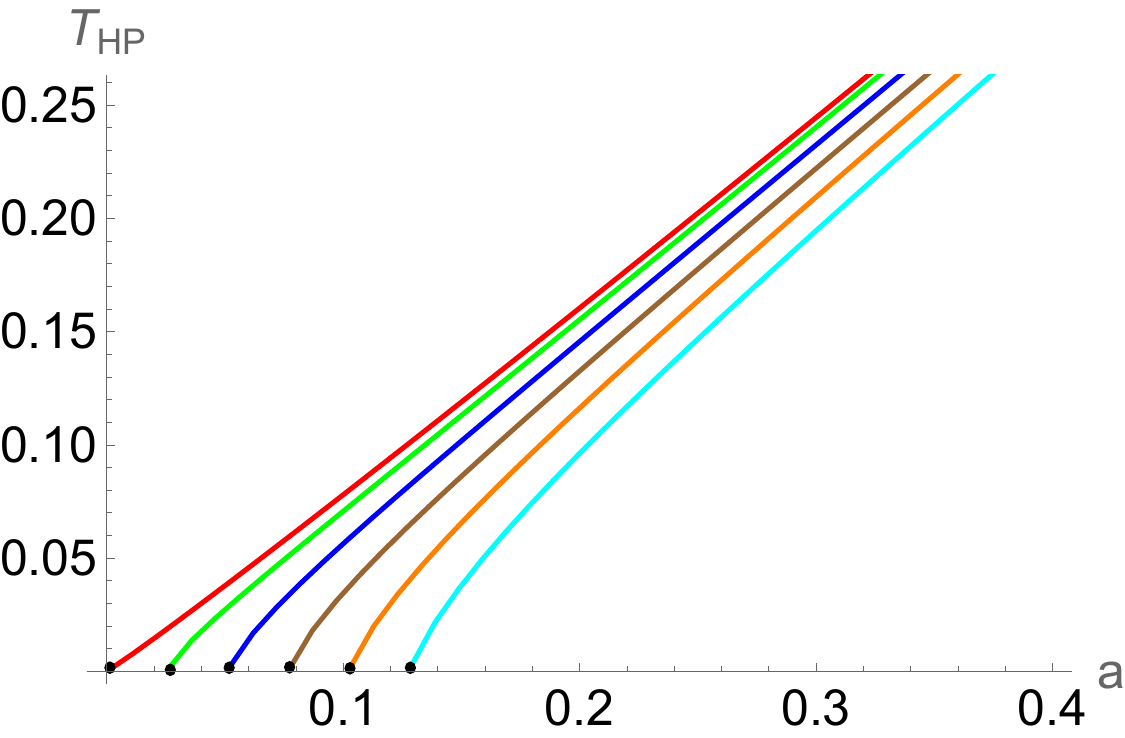}
\caption{ \small Hawking/Page phase transition temperature $T_{HP}$ as a function of $a$ for various values of chemical potential $\mu_{e}$. Red, green, blue, brown, orange, and cyan curves correspond to $\mu_{e}=0$, $0.2$, $0.4$, $0.6$, $0.8$, and $1.0$, respectively. The black dots indicate the critical hairy parameter $a_c$.}
\label{CritTvsaf2}
\end{minipage}
\hspace{0.4cm}
\begin{minipage}[b]{0.5\linewidth}
\centering
\includegraphics[width=2.8in,height=2.3in]{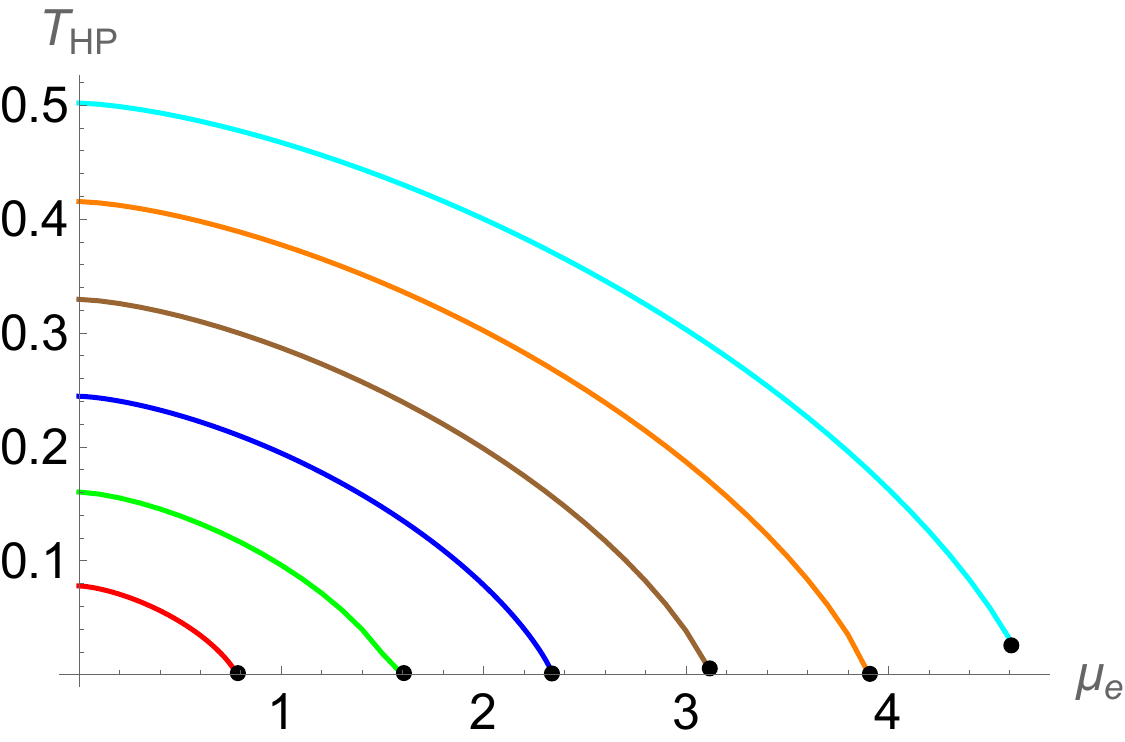}
\caption{\small Hawking/Page phase transition temperature $T_{HP}$ as a function of chemical potential $\mu_{e}$ for various values of $a$. Red, green, blue, brown, orange, and cyan curves correspond to $a=0.1$, $0.2$, $0.3$, $0.4$, $0.5$, and $0.6$, respectively. The black dots indicate the critical chemical potential $\mu_e^{c}$.}
\label{CritTvsmuf2}
\end{minipage}
\end{figure}

\begin{figure}[h!]
\begin{minipage}[b]{0.5\linewidth}
\centering
\includegraphics[width=2.8in,height=2.3in]{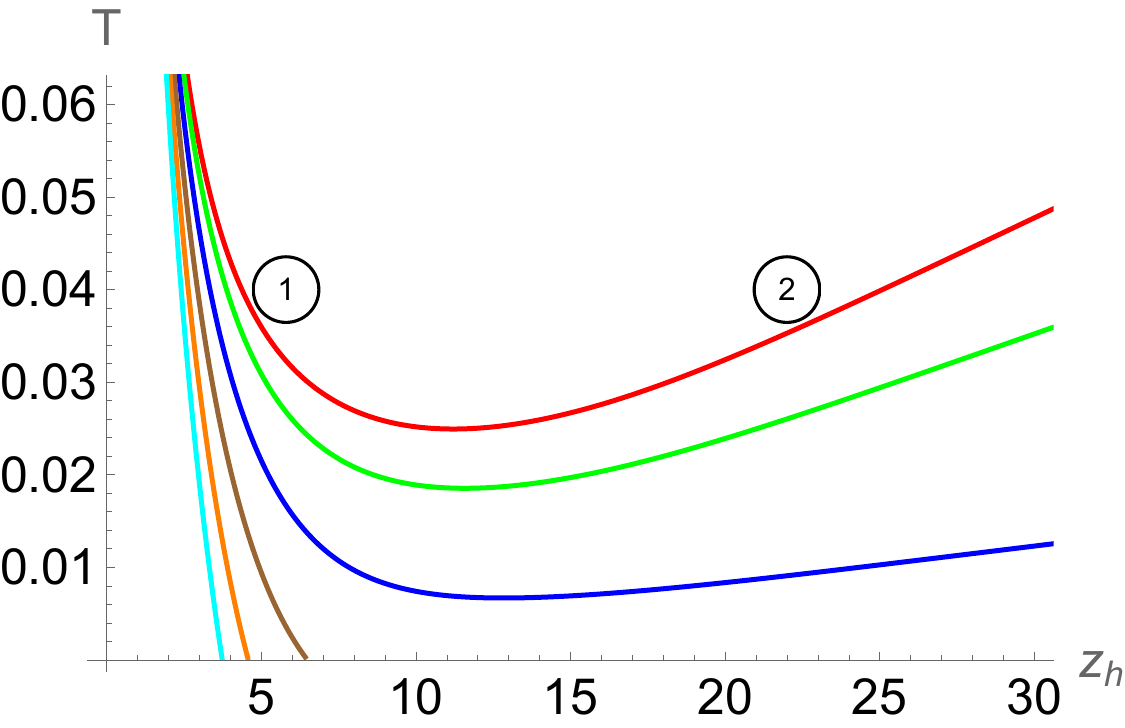}
\caption{ \small Hawking temperature $T$ as a function of horizon radius $z_h$ for various values of charge $q_e$.  Here $a=0.1$ is used. Red, green, blue, brown, orange, and cyan curves correspond to $q_e=0$, $0.1$, $0.2$, $0.3$, $0.4$, and $0.5$, respectively.}
\label{zhvsTvsqeaPt1f2}
\end{minipage}
\hspace{0.4cm}
\begin{minipage}[b]{0.5\linewidth}
\centering
\includegraphics[width=2.8in,height=2.3in]{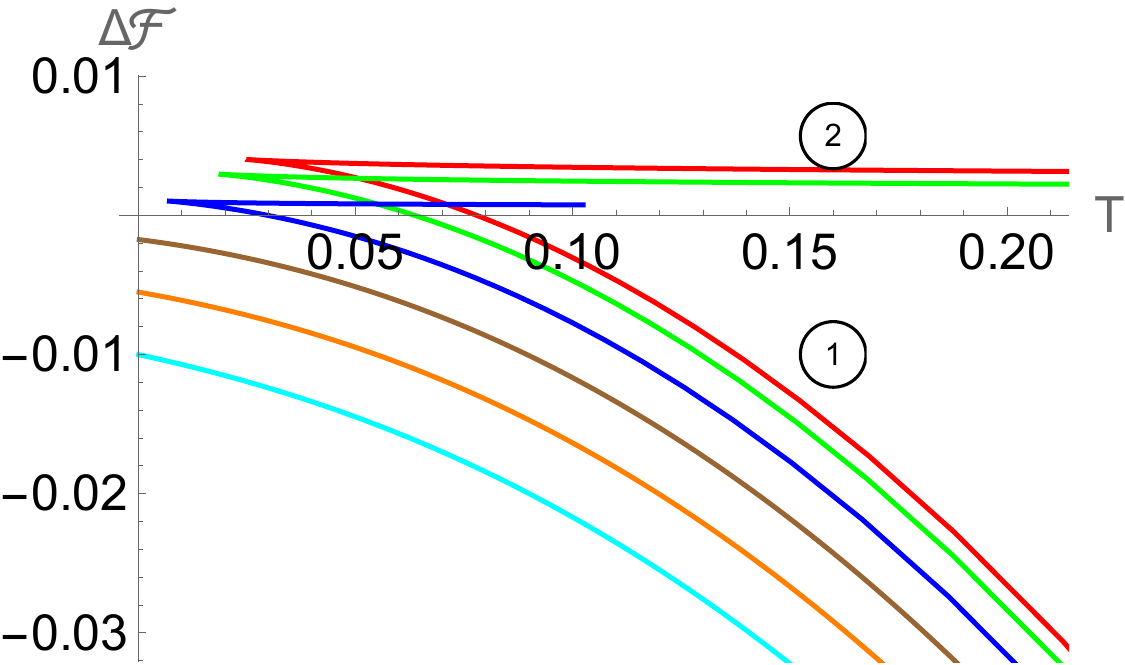}
\caption{\small The Helmholtz free energy difference $\Delta\mathcal{F}$ as a function of $T$ for various values of charge $q_e$. Here, $a=0.1$ is used. Red, green, blue, brown, orange, and cyan curves correspond to $q_e=0$, $0.1$, $0.2$, $0.3$, $0.4$, and $0.5$, respectively.}
\label{TvsFvsqeaPt1f2}
\end{minipage}
\end{figure}

\begin{figure}[h!]
\begin{minipage}[b]{0.5\linewidth}
\centering
\includegraphics[width=2.8in,height=2.3in]{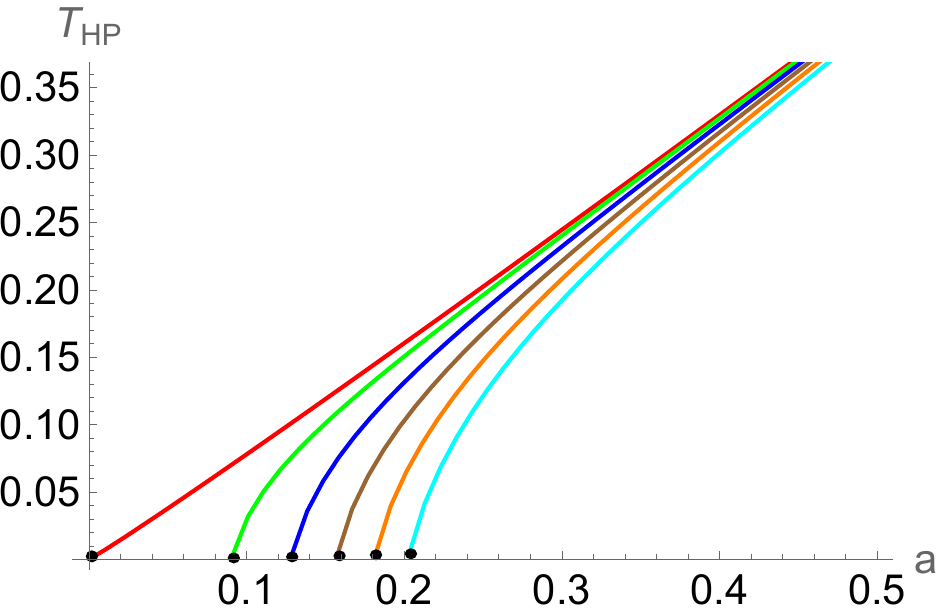}
\caption{ \small Hawking/Page phase transition temperature $T_{HP}$ as a function of $a$ for various values of charge $q_{e}$. Red, green, blue, brown, orange, and cyan curves correspond to $q_{e}=0$, $0.2$, $0.4$, $0.6$, $0.8$, and $1.0$, respectively. The black dots indicate the critical hairy parameter $a_c$.}
\label{CritTvsaf2fixedq}
\end{minipage}
\hspace{0.4cm}
\begin{minipage}[b]{0.5\linewidth}
\centering
\includegraphics[width=2.8in,height=2.3in]{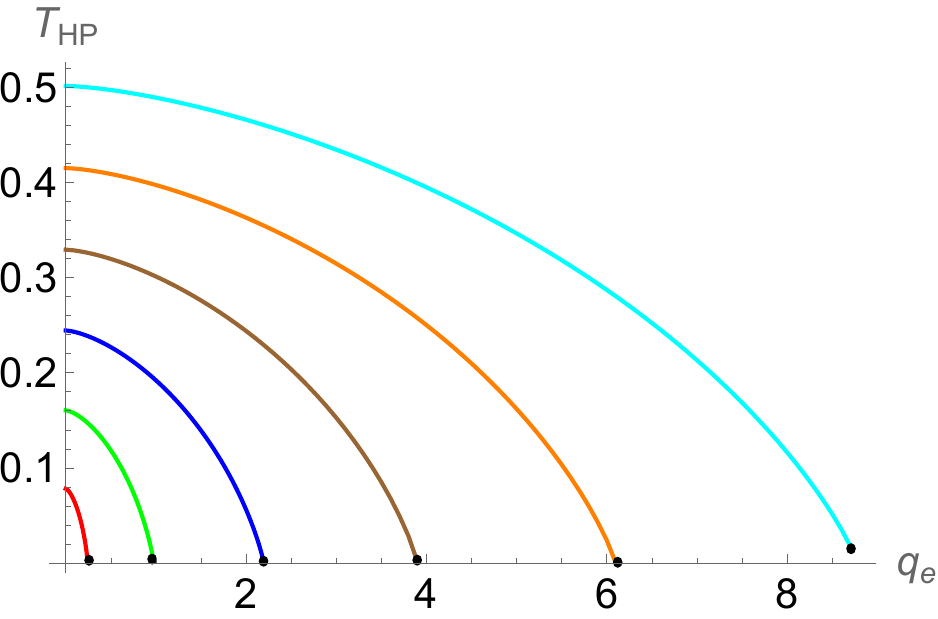}
\caption{\small Hawking/Page phase transition temperature $T_{HP}$ as a function of charge $q_{e}$ for various values of $a$. Red, green, blue, brown, orange, and cyan curves correspond to $a=0.1$, $0.2$, $0.3$, $0.4$, $0.5$, and $0.6$, respectively. The black dots indicate the critical charge $q_{e}^{c}$.}
\label{CritTvsqf2}
\end{minipage}
\end{figure}

The thermodynamic structure with $f(\phi) = e^{\phi}$ coupling remains quite similar to the $f(\phi) = 1$ coupling for the finite chemical potential as well. The results are shown in Figs.~\ref{zhvsTvsmuaPt1f2} and \ref{TvsGvsmuaPt1f2}. Here again, there exists a critical chemical potential $\mu_{e}^{c}$ below which the Hawking/Page phase transition between the thermal-AdS and hairy black hole phases take place, whereas no such phase transition appears above $\mu_{e}^{c}$. In particular, for $\mu_e<\mu_{e}^{c}$, two black hole branches appear which exist only above a certain minimum temperature, whereas, for $\mu_e>\mu_{e}^{c}$, only one black hole branch appears which becomes extremal and remain thermodynamically stable at all temperatures. The magnitudes of $\mu_{e}^{c}$ and the Hawking/Page phase transition $T_{HP}$ again depend nontrivially on the hair parameter $a$. This dependence is shown in Figs.~\ref{CritTvsaf2} and \ref{CritTvsmuf2}. The overall behaviour of $T_{HP}$ concerning $a$ and $\mu_e$ is quite similar to the case of $f(\phi)=1$, albeit with a different magnitude.

We have similarly analyzed the thermodynamic structure in the canonical ensemble. The temperature and free energy profiles are shown in Figs.~\ref{zhvsTvsqeaPt1f2} and \ref{TvsFvsqeaPt1f2}. Again, for small values of $q_e$, two hairy black hole phases appear, with the large stable hairy black hole phase undergoing phase transition to the thermal-AdS phase as the temperature is lowered. Whereas, for relatively large values of $q_e$, only one stable hairy black hole phase appears which becomes extremal at a certain horizon radius. These results again suggest that, just like in the case of $f(\phi)=1$, the charged black hole can undergo a Hawking/Page phase transition depending upon the relative magnitude of $a$ and $q_e$. The structure of $T_{HP}$ again shows monotonic behaviour with $a$ and $q_e$. In particular, $T_{HP}$ increases with $a$ for a fixed $q_e$, whereas it decreases with $q_e$ for a fixed $a$. This is shown in Figs.~\ref{CritTvsaf2fixedq} and \ref{CritTvsqf2}. These results in the canonical ensemble are again quite similar to the $f(\phi)=1$ coupling, albeit with different magnitudes of $T_{HP}$ and critical values $a_c$ and $q_e^c$.

\section{\label{sec:level3} Hairy Black hole solution with Coupling $f(\phi) = e^{\phi^2/2}$}
Now we consider the coupling function $f(\phi) = e^{\phi^2/2}$. Such a coupling function has been thoroughly considered in the hairy black hole context in recent years, see for instance \cite{Herdeiro:2018wub}. Therefore, it is instructive to analyse such coupling functions here as well. With $f(\phi)=e^{\phi^2/2}$ coupling, most of our results for the hairy black hole solution remain the same as in the case of previous coupling $f(\phi)=e^{\phi}$. We will, therefore, be brief here. Since the form factor is the same, the solution for the scalar field will remain the same. This implies that the scalar field continues to be regular, finite, and well-behaved everywhere in the exterior horizon region for this exponential coupling function as well. The solution of the gauge field is now given by
\begin{equation}
B_{t}(z)= \frac{\sqrt{\pi} q_e \left(\text{erf}\left(\sqrt{5} a
   z_h\right)-\text{erf}\left(\sqrt{5} a z\right)\right)}{\sqrt{10} a} \,,
\label{gaugefield3}
\end{equation}
with relation between $\mu_e$ and $q_e$ as
\begin{equation}
q_e= \sqrt{\frac{10}{\pi }} \frac{a \mu _e}{\text{erf}\left(\sqrt{5} a z_h\right)}\,,
\label{gaugefield31}
\end{equation}
which has the same $a\to 0$ limit as with the previous coupling functions. The expression for $g(z)$ comes out to be:
\begin{eqnarray}
& & g(z)  =  \frac{1-e^{a^2 \left(z_h^2-z^2\right)}}{1-e^{a^2 z_h^2}} -\frac{q_e^{3/2} \left(3 \sqrt{10 \pi } e^{-a^2 z^2} \text{erf}\left(\sqrt{5}
   a z\right)-5 \sqrt{3 \pi } \text{erf}\left(\sqrt{6} a z\right)\right)}{80
   a^3} \nonumber\\
& &   + \frac{q_e^{3/2} \sqrt{\pi } e^{-a^2 z^2} \left(e^{a^2 z^2}-1\right) \left(3
   \sqrt{10} \text{erf}\left(\sqrt{5} a z_h\right)-5 \sqrt{3} e^{a^2 z_h^2}
   \text{erf}\left(\sqrt{6} a z_h\right)\right)}{80 a^3 \left(e^{a^2
   z_h^2}-1\right)} \,,
\label{g(z)3}
\end{eqnarray}
which reduces to Eq.~(\ref{gza0limit}) in the $a\to 0$ limit as in the case of previous coupling functions. Therefore, for all different coupling functions considered here, the hairy black hole expressions of various quantities reduce smoothly to the nonhairy expressions in the limit $a\to 0$.

\begin{figure}[ht]
\subfigure[]{
\includegraphics[scale=0.4]{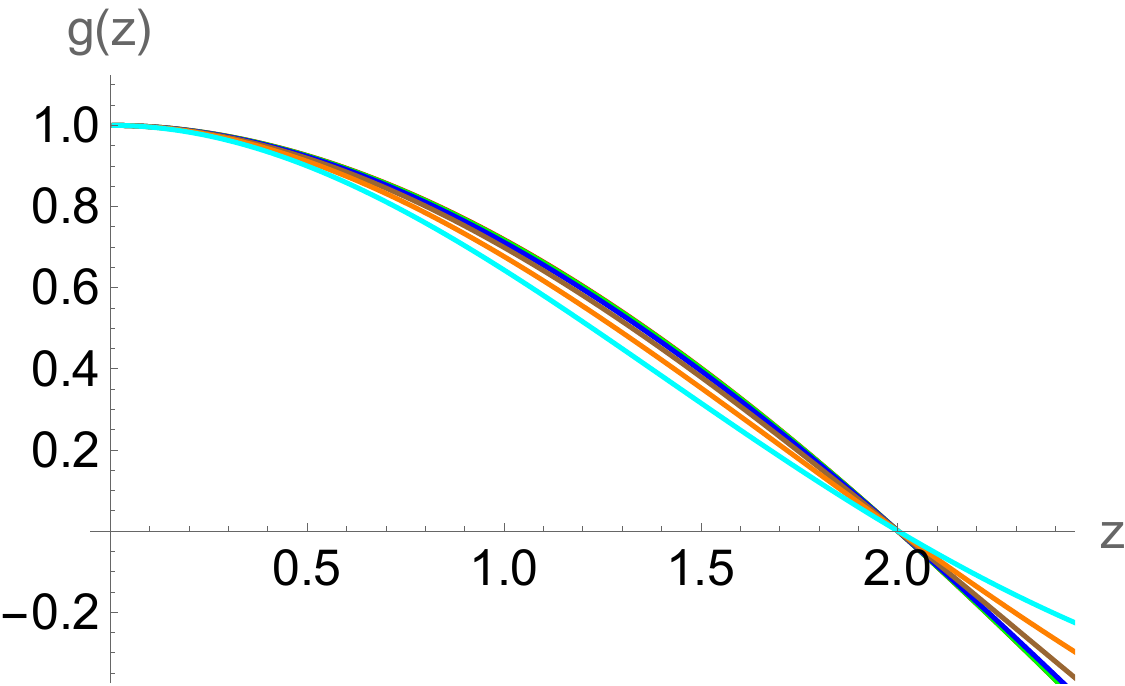}
}
\subfigure[]{
\includegraphics[scale=0.4]{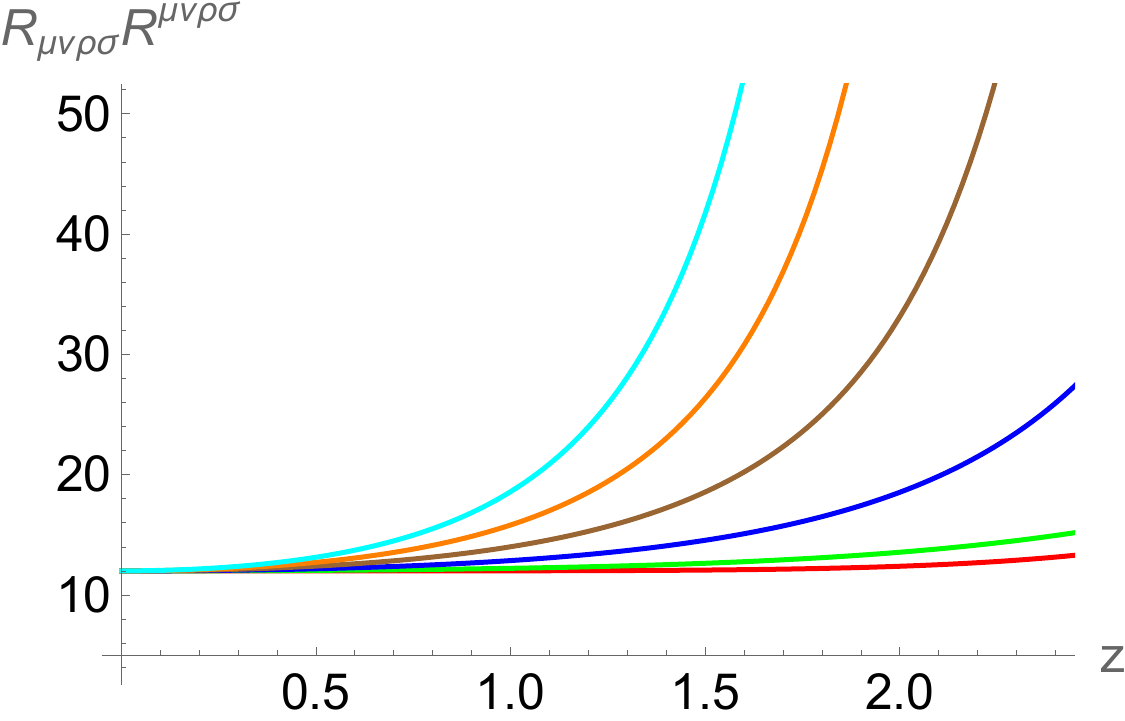}
\label{zvsKretschmannvsazh3qPt2case3}
}
\caption{The nature of $g(z)$ and $R_{\mu\nu\rho\sigma}R^{\mu\nu\rho\sigma}$ for different values of hairy parameter $a$. Here $z_{h}=2.0$ and $q_e=0.2$ are used. Red, green, blue, brown, orange, and cyan curves correspond to $a = 0$, $0.1$, $0.2$, $0.3$, $0.4$, and 0.5, respectively.}
    \label{gvszcase3}
 \end{figure}

The behaviour of $g(z)$ and the Kretschmann scalar for the coupling $f(\phi) = e^{\phi^2/2}$ is shown in Fig.~\ref{gvszcase3}. The hairy black hole solution is again regular and well-behaved everywhere outside the horizon. The curvature is finite everywhere outside the horizon and the singularity appears only inside the horizon. The potential similarly asymptotes to a constant value $V(z)|_{z\rightarrow 0} = 2\Lambda$ at the AdS boundary and is bounded from above. These results firmly establish the well-behaved geometric nature of the hairy black hole with $f(\phi) = e^{\phi^2/2}$ coupling as well.

Let us now briefly talk about the thermodynamics of this black hole. The temperature now has the expression,
\begin{eqnarray}
T & = & \frac{a^2 z_h
   e^{a^2 z_h^2}}{2 \pi  \left(e^{a^2 z_h^2}-1\right)} + \frac{q_e^{3/2} z_h e^{a^2 z_h^2} \left(5 \sqrt{3 \pi }
   \text{erf}\left(\sqrt{6} a z_h\right)-3 \sqrt{10 \pi }
   \text{erf}\left(\sqrt{5} a z_h\right)\right)}{160 \pi  a \left(e^{a^2
   z_h^2}-1\right)} \,,
\label{tempcase3}
\end{eqnarray}
which also reduces to Eq.~(\ref{tempcase1a0}) in the limit $a\rightarrow 0$ and to the standard BTZ expression in the limits $a\rightarrow 0$ and $q_e\rightarrow 0$.

\begin{figure}[h!]
\begin{minipage}[b]{0.5\linewidth}
\centering
\includegraphics[width=2.8in,height=2.3in]{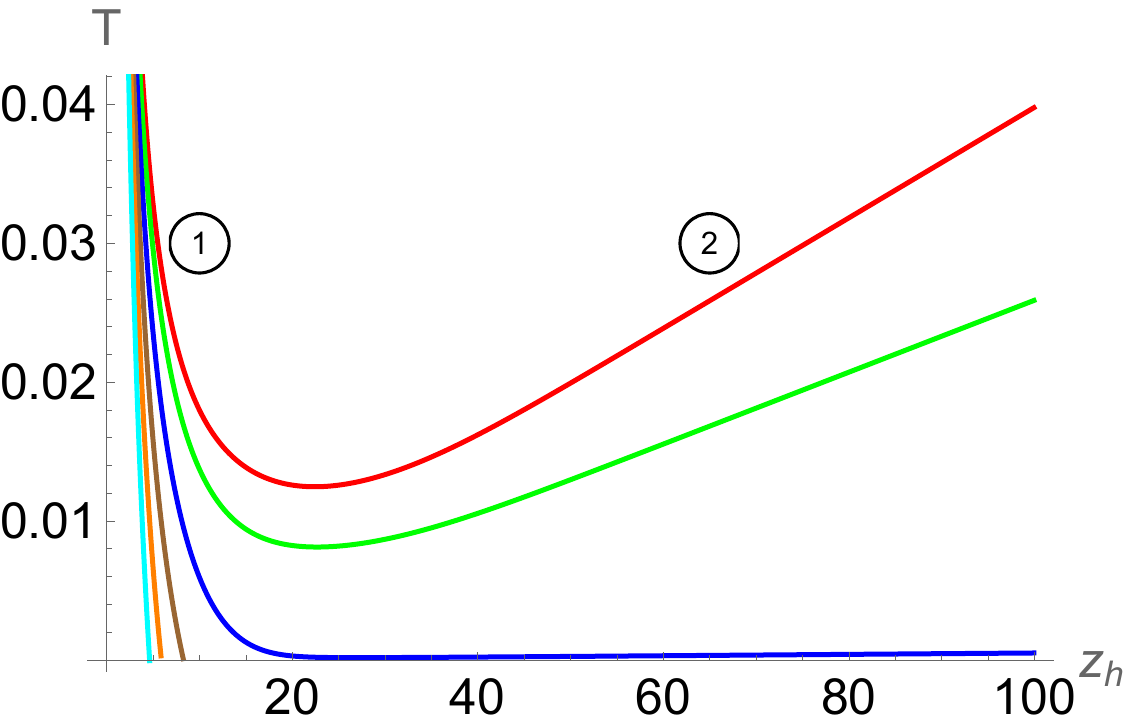}
\caption{ \small Hawking temperature $T$ as a function of horizon radius $z_h$ for various values of chemical potential $\mu_e$. Here $a=0.05$ is used. Red, green, blue, brown, orange, and cyan curves correspond to $\mu_e=0$, $0.2$, $0.4$, $0.6$, $0.8$, and $1.0$, respectively.}
\label{zhvsTvsmuaPt05f3}
\end{minipage}
\hspace{0.4cm}
\begin{minipage}[b]{0.5\linewidth}
\centering
\includegraphics[width=2.8in,height=2.3in]{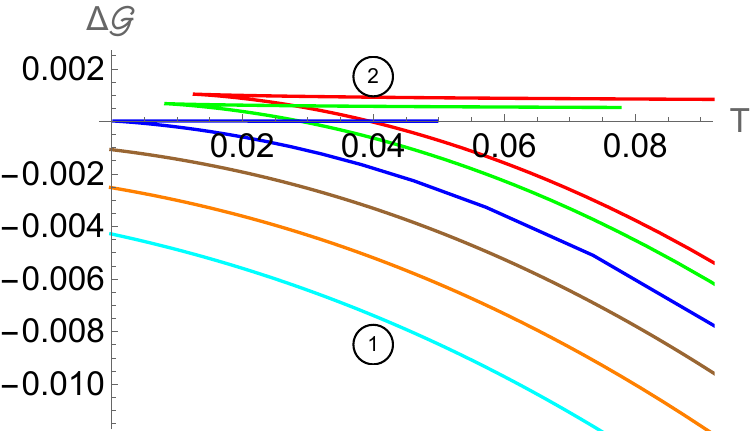}
\caption{\small The Gibbs free energy difference $\Delta\mathcal{G}$ as a function of $T$ for various values of chemical potential $\mu_e$. Here $a=0.05$ is used. Red, green, blue, brown, orange, and cyan curves correspond to $\mu_e=0$, $0.2$, $0.4$, $0.6$, $0.8$, and $1.0$, respectively.}
\label{TvsGvsmuaPt05f3}
\end{minipage}
\end{figure}
\begin{figure}[h!]
\begin{minipage}[b]{0.5\linewidth}
\centering
\includegraphics[width=2.8in,height=2.3in]{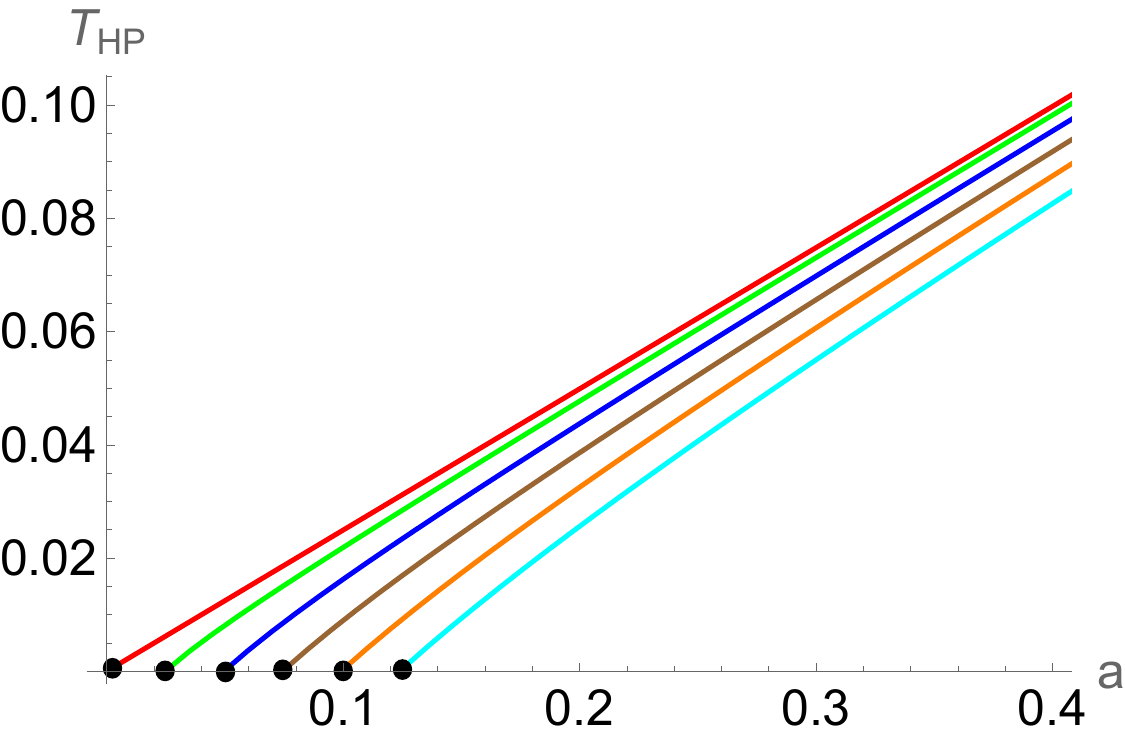}
\caption{ \small Hawking/Page phase transition temperature $T_{HP}$ as a function of $a$ for various values of chemical potential $\mu_{e}$. Red, green, blue, brown, orange, and cyan curves correspond to $\mu_{e}=0$, $0.2$, $0.4$, $0.6$, $0.8$, and $1.0$, respectively. The black dots indicate the
critical hairy parameter $a_c$.}
\label{CritTvsaf3}
\end{minipage}
\hspace{0.4cm}
\begin{minipage}[b]{0.5\linewidth}
\centering
\includegraphics[width=2.8in,height=2.3in]{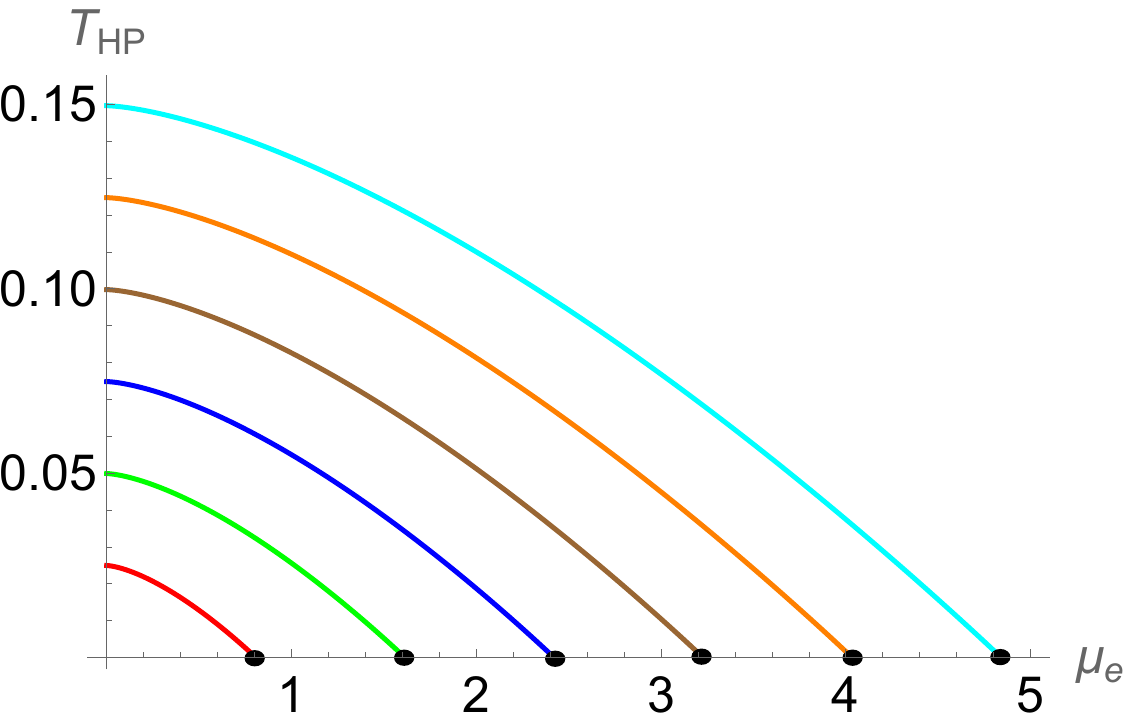}
\caption{\small Hawking/Page phase transition temperature $T_{HP}$ as a function of chemical potential $\mu_{e}$ for various values of $a$. Red, green, blue, brown, orange, and cyan curves correspond to $a=0.1$, $0.2$, $0.3$, $0.4$, $0.5$, and $0.6$, respectively. The black dots indicate the
critical chemical potential $\mu_{e}^{c}$.}
\label{CritTvsmuf3}
\end{minipage}
\end{figure}

\begin{figure}[h!]
\begin{minipage}[b]{0.5\linewidth}
\centering
\includegraphics[width=2.8in,height=2.3in]{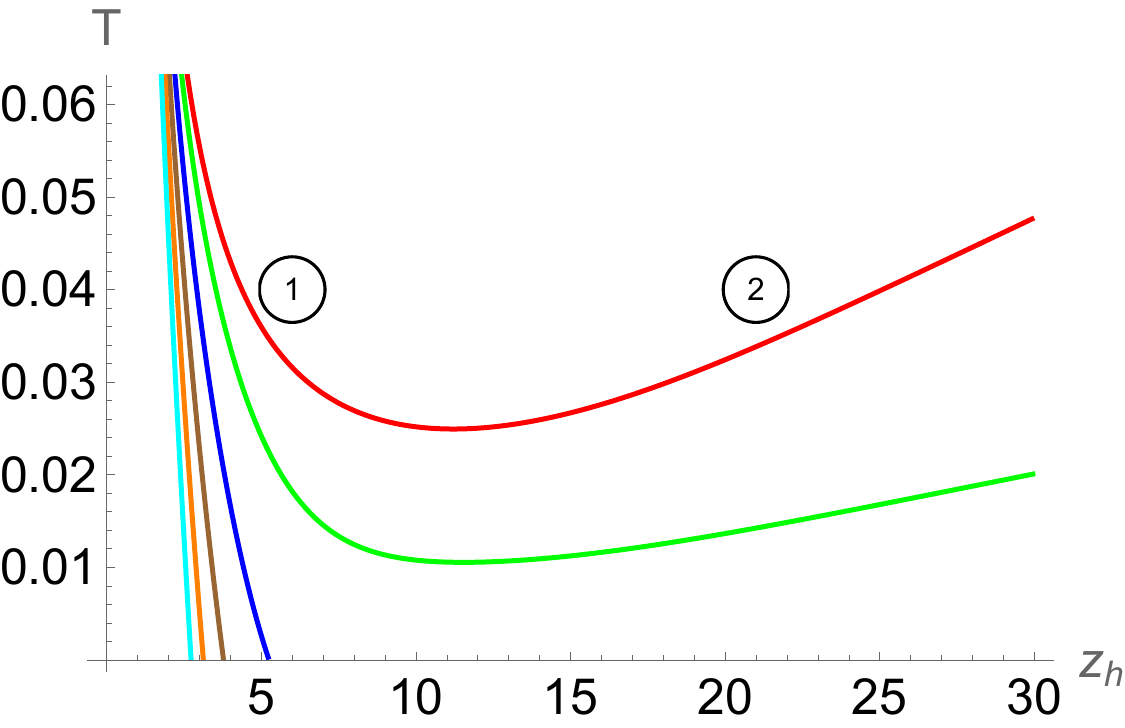}
\caption{ \small Hawking temperature $T$ as a function of horizon radius $z_h$ for various values of charge $q_e$.  Here, $a=0.1$ is used. Red, green, blue, brown, orange, and cyan curves correspond to $q_e=0$, $0.1$, $0.2$, $0.3$, $0.4$, and $0.5$, respectively.}
\label{zhvsTvsqeaPt1f3}
\end{minipage}
\hspace{0.4cm}
\begin{minipage}[b]{0.5\linewidth}
\centering
\includegraphics[width=2.8in,height=2.3in]{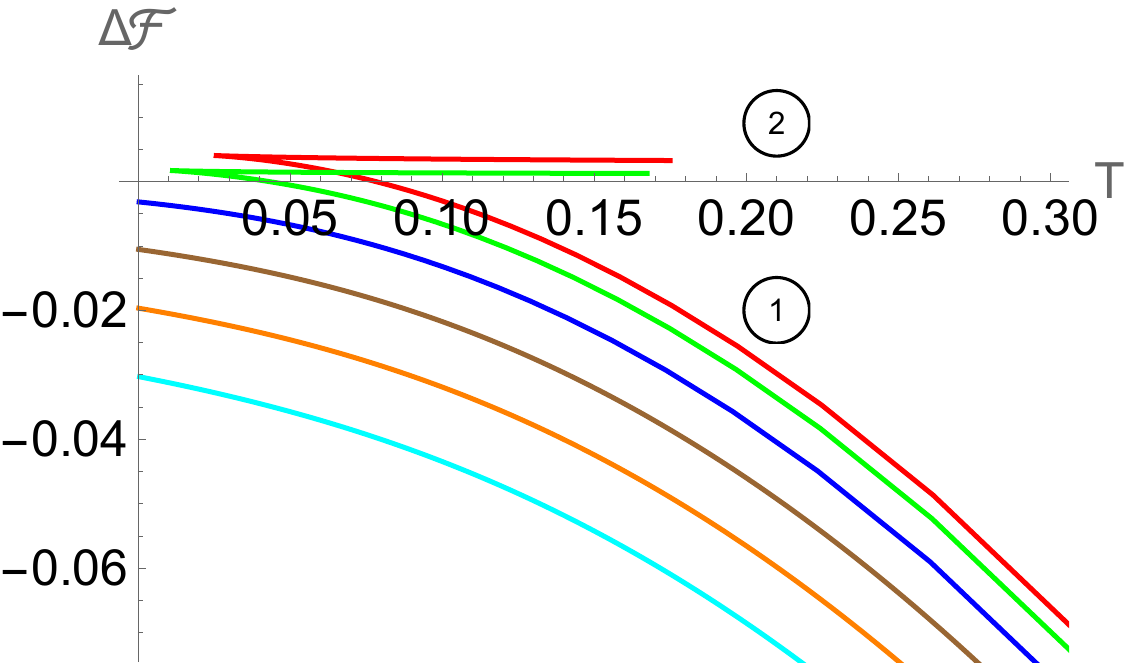}
\caption{\small The Helmholtz free energy difference $\Delta\mathcal{F}$ as a function of $T$ for various values of charge $q_e$. Here, $a=0.1$ is used. Red, green, blue, brown, orange, and cyan curves correspond to $q_e=0$, $0.1$, $0.2$, $0.3$, $0.4$, and $0.5$, respectively.}
\label{TvsFvsqeaPt1f3}
\end{minipage}
\end{figure}

\begin{figure}[h!]
\begin{minipage}[b]{0.5\linewidth}
\centering
\includegraphics[width=2.8in,height=2.3in]{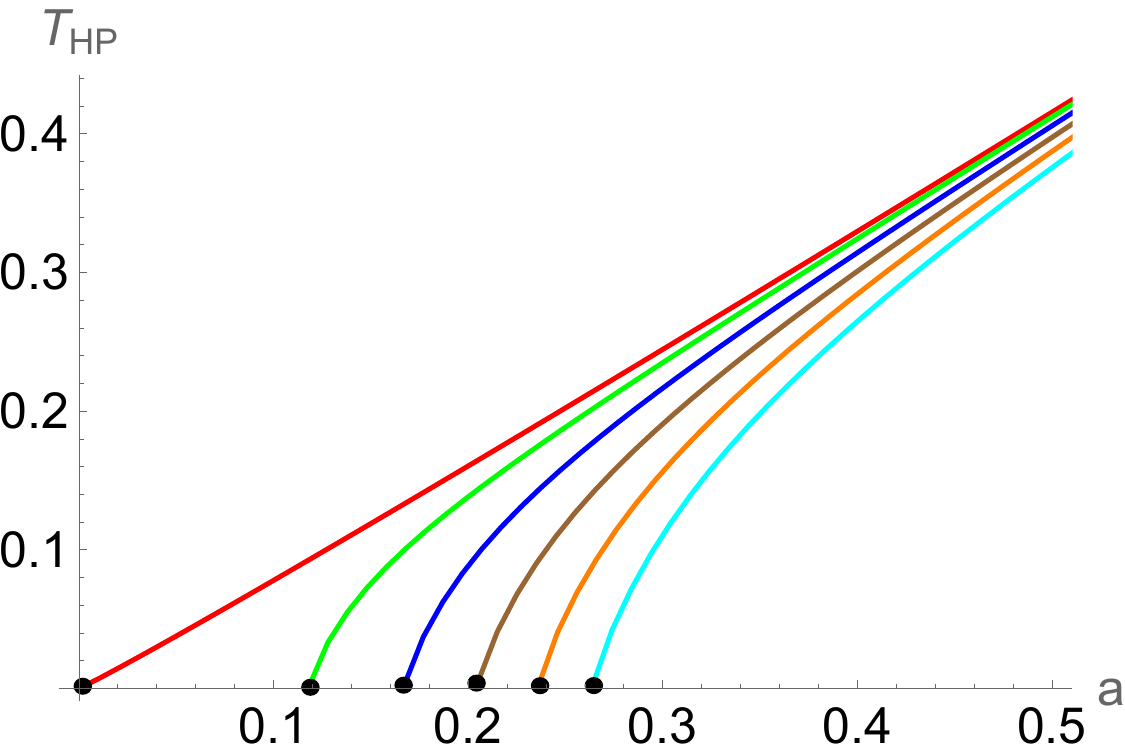}
\caption{ \small Hawking/Page phase transition temperature $T_{HP}$ as a function of $a$ for various values of charge $q_{e}$. Red, green, blue, brown, orange, and cyan curves correspond to $q_{e}=0$, $0.2$, $0.4$, $0.6$, $0.8$, and $1.0$, respectively. The black dots indicate the
critical hairy parameter $a_c$.}
\label{CritTvsaf3fixedq}
\end{minipage}
\hspace{0.4cm}
\begin{minipage}[b]{0.5\linewidth}
\centering
\includegraphics[width=2.8in,height=2.3in]{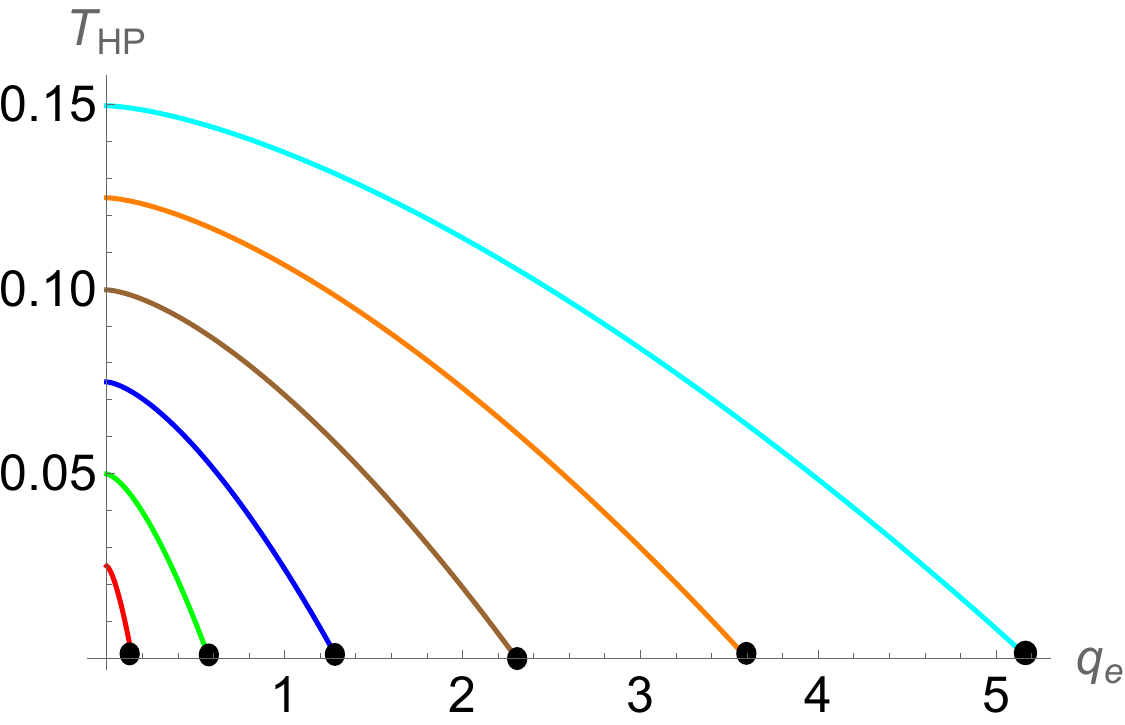}
\caption{\small Hawking/Page phase transition temperature $T_{HP}$ as a function of charge $q_{e}$ for various values of $a$. Red, green, blue, brown, orange, and cyan curves correspond to $a=0.1$, $0.2$, $0.3$, $0.4$, $0.5$, and $0.6$, respectively. The black dots indicate the
critical charge $q_{e}^{c}$.}
\label{CritTvsqf3}
\end{minipage}
\end{figure}

The thermodynamic structure of the hairy black hole in the grand-canonical ensemble is shown in Figs.~\ref{zhvsTvsmuaPt05f3} and \ref{TvsGvsmuaPt05f3}. Since for $\mu_e=0$, the Einstein-power Maxwell-scalar gravity action becomes identical for all $f(\phi)$, it ensures identical thermodynamic structure for hairy black holes at $\mu_e=0$ for all $f(\phi)$. For $f(\phi) = e^{\phi^2/2}$ and for small chemical potential, there exists a thermodynamically stable hairy black hole phase which undergoes a phase transition to thermal-AdS phase as the temperature is lowered, i.e., the Hawking/page phase transition continues to exist, with the thermal-AdS phase dominating the structure at lower temperatures. In contrast, a large stable hairy black hole phase dominates the phase structure at higher temperatures. Similarly, there exists a critical chemical potential $\mu_{e}^{c}$ above which the Hawking/Page phase transition ceases to exist. Therefore, here again for $\mu_e<\mu_{e}^{c}$, two black hole branches appear which exist only above a certain minimum temperature, whereas, for $\mu_e>\mu_{e}^{c}$, only one black hole branch appears which can become extremal and remain thermodynamically stable at all temperatures. The thermodynamically favoured black holes are also thermodynamically stable as they have positive specific heat. The magnitudes of $\mu_{e}^{c}$ and the Hawking/Page phase transition $T_{HP}$ again depend nontrivially on the hair parameter $a$. This dependence is shown in Figs.~\ref{CritTvsaf3} and \ref{CritTvsmuf3}.

The thermodynamic results in the canonical ensemble are shown in Figs.~\ref{zhvsTvsqeaPt1f3} and \ref{TvsFvsqeaPt1f3}. Here again, we find that depending upon the relative magnitude of $a$ and $q_e$, the fixed-charged hairy black hole undergoes a Hawking/Page phase transition as the temperature is varied. The corresponding phase transition temperature and critical point $q_e^c$ behaviour are illustrated in Figs.~\ref{CritTvsaf3fixedq} and \ref{CritTvsqf3}.

We end this section by emphasizing that with $f(\phi) = e^{\phi^2/2}$ coupling, the thermodynamic phase diagram of the hairy black hole, both in the canonical and grand-canonical ensemble, remains quite similar to the $f(\phi) = 1$ and $f(\phi) = e^{\phi}$ cases. Our investigation, therefore, indicates some type of universality in the thermodynamic phase structure of the hairy black hole with Coulomb-like potential for different coupling functions. Particularly, there exist critical points $\{ \mu_e^c, q_e^c\}$, the magnitude of which are coupling function dependent, below which there appears first-order phase transition between the large hairy black hole and thermal-AdS phases, whereas above these critical points, no such phase transition exits. We analyzed some other forms of $A(z)$ as well and found similar universal results in the thermodynamic phase structure of hairy black holes.

\section{\label{sec:level1}  Conclusions}
In this paper, we have constructed a new family of three-dimensional hairy-charged black hole solutions from the Einstein-power Maxwell-scalar action. The gauge field solution in particular, being devoid of logarithmic singularity, is everywhere well-behaved and reduces to the usual inverse power law behavior for the non-hairy case, thereby downplaying the usual issues faced in Maxwell electrodynamic in three dimensions.  The constructed solutions were based on two functions: the coupling function $f(\phi)$ and the form factor $A(z)$. We specifically analysed the solutions for three interesting and physically motivated forms of the coupling function: (i) $f(\phi)=1$, (ii) $f(\phi)=e^{\phi}$, and lastly (iii) $f(\phi)=e^{\phi^2/2}$, along with the simple form of $A(z)=-a^2z^2$. The parameter $a$ regulates the strength of the scalar hair, and in the limit $a\to 0$, the solution always reduces to the standard non-hairy BTZ black hole with a Coulomb-like potential. We have observed that in each of the solutions: (i) the scalar hair is found to be regular everywhere outside the horizon and goes to zero at the asymptotic AdS boundary, (ii) the Kretschmann and Ricci scalars are always finite and well-behaved outside the horizon, and diverge only at the centre of the black hole, and (iii) the potential is found to be bounded from above from its UV boundary value. These results indicate the smooth and desirable nature of the constructed hairy black holes for these different coupling functions.

Next, we analysed the thermodynamic properties of the hairy black hole solutions in canonical and grand-canonical ensembles and found some universal and intriguing results. For each of the considered coupling functions, a critical value of the hairy parameter $a_{c}$ appeared above which the black hole exhibited the Hawking/Page phase transition to the thermal-AdS phase as the temperature is lowered, whereas below this $a_{c}$ no phase transition existed. This result should be contrasted with the usual uncharged BTZ black hole case, where no such phase transition appeared. This suggests that the addition of a scalar hair makes the three-dimensional phase structure much richer. Additionally, it is observed that the associated transition temperature also increases monotonically with $a$. We similarly analysed the hairy thermodynamic structure for finite values of $q_e$ and $\mu_e$ in the canonical and grand-canonical ensembles and found that the Hawking/Page phase transition continues to persist for small values of $q_e$ and $\mu_e$, whereas for large values of $\mu_e$ and $q_e$ no such phase transition occurs. These results indicate the existence of critical values $q_{e}^{c}$ and $\mu_{e}^{c}$ (which are $f(\phi)$ dependent) at which the Hawking/Page phase transition line stops. The transition temperature was further found to be decreasing monotonically with $q_e$ and $\mu_e$. Interestingly, this thermodynamic pattern matches quite well with the charged RN-AdS black holes in four and higher dimensions. This is intriguing because while BTZ black holes and their counterparts in higher dimensions share a number of geometric characteristics, their thermodynamic structures are very different.  We also found that the specific heat is always positive in the thermodynamically favoured hairy black hole phase, hence establishing the local stability of the hairy black holes.

This work might be extended in many directions. It would
be interesting to extend this work by finding its axisymmetric counterpart. We anticipate that similar to the BTZ black hole, the charged hairy black hole thermodynamic structure may be significantly altered by the rotational parameter. It is also important to check the dynamical stability of the constructed hairy black hole under various perturbations. Our preliminary analysis in this direction leads us to believe that these hairy black holes are dynamically stable under scalar field perturbations. Work in these directions is in progress.

\section*{Acknowledgements}
A.D. would like to express his sincere gratitude to S. Priyadarshinee and S.S. Jena for their insightful discussions throughout the course of this research. The work of SM is supported by the Department of Science and Technology, Government of India under the Grant Agreement number IFA17-PH207 (INSPIRE Faculty Award).

\bibliography{btz}

\begin{thebibliography}{100}

\bibitem{Hawking:1975vcx}
S.~W. Hawking.
\newblock {Particle Creation by Black Holes}.
\newblock {\em Commun. Math. Phys.}, 43:199--220, 1975.
\newblock [Erratum: Commun.Math.Phys. 46, 206 (1976)].

\bibitem{Gibbons:1976ue}
G.~W. Gibbons and S.~W. Hawking.
\newblock {Action Integrals and Partition Functions in Quantum Gravity}.
\newblock {\em Phys. Rev. D}, 15:2752--2756, 1977.

\bibitem{PhysRevD.7.2333}
Jacob~D. Bekenstein.
\newblock Black holes and entropy.
\newblock {\em Phys. Rev. D}, 7:2333--2346, Apr 1973.

\bibitem{Hawking:1982dh}
S.~W. Hawking and Don~N. Page.
\newblock {Thermodynamics of Black Holes in anti-De Sitter Space}.
\newblock {\em Commun. Math. Phys.}, 87:577, 1983.

\bibitem{PhysRevD.60.064018}
Andrew Chamblin, Roberto Emparan, Clifford~V. Johnson, and Robert~C. Myers.
\newblock Charged ads black holes and catastrophic holography.
\newblock {\em Phys. Rev. D}, 60:064018, Aug 1999.

\bibitem{PhysRevD.60.104026}
Andrew Chamblin, Roberto Emparan, Clifford~V. Johnson, and Robert~C. Myers.
\newblock Holography, thermodynamics, and fluctuations of charged ads black
  holes.
\newblock {\em Phys. Rev. D}, 60:104026, Oct 1999.

\bibitem{Cvetic:1999ne}
Mirjam Cvetic and Steven~S. Gubser.
\newblock {Phases of R charged black holes, spinning branes and strongly
  coupled gauge theories}.
\newblock {\em JHEP}, 04:024, 1999.

\bibitem{Sahay:2010tx}
Anurag Sahay, Tapobrata Sarkar, and Gautam Sengupta.
\newblock {On the Thermodynamic Geometry and Critical Phenomena of AdS Black
  Holes}.
\newblock {\em JHEP}, 07:082, 2010.

\bibitem{Sahay:2010wi}
Anurag Sahay, Tapobrata Sarkar, and Gautam Sengupta.
\newblock {Thermodynamic Geometry and Phase Transitions in Kerr-Newman-AdS
  Black Holes}.
\newblock {\em JHEP}, 04:118, 2010.

\bibitem{Dey:2015ytd}
Anshuman Dey, Subhash Mahapatra, and Tapobrata Sarkar.
\newblock {Thermodynamics and Entanglement Entropy with Weyl Corrections}.
\newblock {\em Phys. Rev. D}, 94(2):026006, 2016.

\bibitem{Mahapatra:2016dae}
Subhash Mahapatra.
\newblock {Thermodynamics, Phase Transition and Quasinormal modes with Weyl
  corrections}.
\newblock {\em JHEP}, 04:142, 2016.

\bibitem{Banados:1992wn}
Maximo Banados, Claudio Teitelboim, and Jorge Zanelli.
\newblock {The Black hole in three-dimensional space-time}.
\newblock {\em Phys. Rev. Lett.}, 69:1849--1851, 1992.

\bibitem{PhysRevD.48.1506}
M\'aximo Ba\~nados, Marc Henneaux, Claudio Teitelboim, and Jorge Zanelli.
\newblock Geometry of the 2+1 black hole.
\newblock {\em Phys. Rev. D}, 48:1506--1525, Aug 1993.

\bibitem{Brown:1986nw}
J.~David Brown and M.~Henneaux.
\newblock {Central Charges in the Canonical Realization of Asymptotic
  Symmetries: An Example from Three-Dimensional Gravity}.
\newblock {\em Commun. Math. Phys.}, 104:207--226, 1986.

\bibitem{Maldacena:1997re}
Juan~Martin Maldacena.
\newblock {The Large N limit of superconformal field theories and
  supergravity}.
\newblock {\em Adv. Theor. Math. Phys.}, 2:231--252, 1998.

\bibitem{Strominger:1997eq}
Andrew Strominger.
\newblock {Black hole entropy from near horizon microstates}.
\newblock {\em JHEP}, 02:009, 1998.

\bibitem{Achucarro:1986uwr}
A.~Achucarro and P.~K. Townsend.
\newblock {A Chern-Simons Action for Three-Dimensional anti-De Sitter
  Supergravity Theories}.
\newblock {\em Phys. Lett. B}, 180:89, 1986.

\bibitem{Witten:1988hc}
Edward Witten.
\newblock {(2+1)-Dimensional Gravity as an Exactly Soluble System}.
\newblock {\em Nucl. Phys. B}, 311:46, 1988.

\bibitem{Carlip:1995qv}
Steven Carlip.
\newblock {The (2+1)-Dimensional black hole}.
\newblock {\em Class. Quant. Grav.}, 12:2853--2880, 1995.

\bibitem{Luongo:2023xaw}
Orlando Luongo, Hernando Quevedo, and S.~N. Sajadi.
\newblock {Gravitational repulsive effects in 3D regular black holes}.
\newblock 11 2023.

\bibitem{Sajadi:2023ybm}
Seyed~Naseh Sajadi, Mohsen Khodadi, Orlando Luongo, and Hernando Quevedo.
\newblock {Anisotropic Generalized Polytropic Spheres: Regular 3D Black Holes}.
\newblock 12 2023.

\bibitem{Martinez:1999qi}
Cristian Martinez, Claudio Teitelboim, and Jorge Zanelli.
\newblock {Charged rotating black hole in three space-time dimensions}.
\newblock {\em Phys. Rev. D}, 61:104013, 2000.

\bibitem{Bergshoeff:2009hq}
Eric~A. Bergshoeff, Olaf Hohm, and Paul~K. Townsend.
\newblock {Massive Gravity in Three Dimensions}.
\newblock {\em Phys. Rev. Lett.}, 102:201301, 2009.

\bibitem{Perez:2013xi}
Alfredo Perez, David Tempo, and Ricardo Troncoso.
\newblock {Higher spin black hole entropy in three dimensions}.
\newblock {\em JHEP}, 04:143, 2013.

\bibitem{Ammon:2011nk}
Martin Ammon, Michael Gutperle, Per Kraus, and Eric Perlmutter.
\newblock {Spacetime Geometry in Higher Spin Gravity}.
\newblock {\em JHEP}, 10:053, 2011.

\bibitem{Cataldo:2000we}
Mauricio Cataldo, Norman Cruz, Sergio del Campo, and Alberto Garcia.
\newblock {(2+1)-dimensional black hole with Coulomb - like field}.
\newblock {\em Phys. Lett. B}, 484:154, 2000.

\bibitem{Garcia-Diaz:2017cpv}
Alberto~A. Garc\'\i{}a-D\'\i{}az.
\newblock {\em {Exact Solutions in Three-Dimensional Gravity}}.
\newblock Cambridge Monographs on Mathematical Physics. Cambridge University
  Press, 9 2017.

\bibitem{Larranaga:2008qw}
Alexis Larranaga and Luz~Angela Garcia.
\newblock {Thermodynamics of the Three-dimensional Black Hole with a
  Coulomb-like Field}.
\newblock {\em Electron. J. Theor. Phys.}, 9(27):121--130, 2012.

\bibitem{Larranaga:2008fn}
Alexis Larranaga.
\newblock {Thermodynamics of the (2+1)-dimensional Black Hole with non linear
  Electrodynamics and without Cosmological Constant from the Generalized
  Uncertainly Principle}.
\newblock {\em Bulg. J. Phys.}, 37:10--15, 2010.

\bibitem{Balart:2009et}
Leonardo Balart.
\newblock {Energy distribution of 2+1 dimensional black holes with nonlinear
  electrodynamics}.
\newblock {\em Mod. Phys. Lett. A}, 24:2777--2785, 2009.

\bibitem{Mazharimousavi:2011nd}
S.~Habib Mazharimousavi, O.~Gurtug, M.~Halilsoy, and O.~Unver.
\newblock {2+1 dimensional magnetically charged solutions in Einstein - Power -
  Maxwell theory}.
\newblock {\em Phys. Rev. D}, 84:124021, 2011.

\bibitem{Balart:2019uok}
Leonardo Balart and Sharmanthie Fernando.
\newblock {Non-linear black holes in 2+1 dimensions as heat engines}.
\newblock {\em Phys. Lett. B}, 795:638--643, 2019.

\bibitem{Rincon:2018dsq}
\'Angel Rinc\'on, Ernesto Contreras, Pedro Bargue\~no, Benjamin Koch, and
  Grigorios Panotopoulos.
\newblock {Scale-dependent ( $2+1$ )-dimensional electrically charged black
  holes in Einstein-power-Maxwell theory}.
\newblock {\em Eur. Phys. J. C}, 78(8):641, 2018.

\bibitem{Rincon:2017goj}
\'Angel Rinc\'on, Ernesto Contreras, Pedro Bargue\~no, Benjamin Koch, Grigorios
  Panotopoulos, and Alejandro Hern\'andez-Arboleda.
\newblock {Scale dependent three-dimensional charged black holes in linear and
  non-linear electrodynamics}.
\newblock {\em Eur. Phys. J. C}, 77(7):494, 2017.

\bibitem{Dehghani:2016agl}
M.~Dehghani.
\newblock {Thermodynamics of $(2+1)$-dimensional charged black holes with
  power-law Maxwell field}.
\newblock {\em Phys. Rev. D}, 94(10):104071, 2016.

\bibitem{Cataldo:2020cxm}
Mauricio Cataldo, P.~A. Gonz\'alez, Joel Saavedra, Yerko V\'asquez, and Bin
  Wang.
\newblock {Thermodynamics of ( 2+1 )-dimensional Coulomb-like black holes from
  nonlinear electrodynamics with a traceless energy momentum tensor}.
\newblock {\em Phys. Rev. D}, 103(2):024047, 2021.

\bibitem{HabibMazharimousavi:2013duq}
S.~Habib~Mazharimousavi, M.~Halilsoy, and O.~Gurtug.
\newblock {A new Einstein-nonlinear electrodynamics solution in 2 + 1
  dimensions}.
\newblock {\em Eur. Phys. J. C}, 74(1):2735, 2014.

\bibitem{Gonzalez:2021vwp}
P.~A. Gonz\'alez, \'Angel Rinc\'on, Joel Saavedra, and Yerko V\'asquez.
\newblock {Superradiant instability and charged scalar quasinormal modes for
  (2+1)-dimensional Coulomb-like AdS black holes from nonlinear
  electrodynamics}.
\newblock {\em Phys. Rev. D}, 104(8):084047, 2021.

\bibitem{Amirabi:2021uam}
Z.~Amirabi.
\newblock {Generalized Einstein\textendash{}Power Maxwell theory in
  2+1-dimensions}.
\newblock {\em Eur. Phys. J. Plus}, 136(5):569, 2021.

\bibitem{Aragon:2021ogo}
Almendra Arag\'on, P.~A. Gonz\'alez, Joel Saavedra, and Yerko V\'asquez.
\newblock {Scalar quasinormal modes for $2+1$-dimensional Coulomb-like AdS
  black holes from nonlinear electrodynamics}.
\newblock {\em Gen. Rel. Grav.}, 53(10):91, 2021.

\bibitem{Sheykhi:2012zz}
Ahmad Sheykhi.
\newblock {Higher-dimensional charged $f(R)$ black holes}.
\newblock {\em Phys. Rev. D}, 86:024013, 2012.

\bibitem{Martinez:1996gn}
Cristian Martinez and Jorge Zanelli.
\newblock {Conformally dressed black hole in (2+1)-dimensions}.
\newblock {\em Phys. Rev. D}, 54:3830--3833, 1996.

\bibitem{Henneaux:2002wm}
Marc Henneaux, Cristian Martinez, Ricardo Troncoso, and Jorge Zanelli.
\newblock {Black holes and asymptotics of 2+1 gravity coupled to a scalar
  field}.
\newblock {\em Phys. Rev. D}, 65:104007, 2002.

\bibitem{Chan:1994qa}
K.~C.~K. Chan and Robert~B. Mann.
\newblock {Static charged black holes in (2+1)-dimensional dilaton gravity}.
\newblock {\em Phys. Rev. D}, 50:6385, 1994.
\newblock [Erratum: Phys.Rev.D 52, 2600 (1995)].

\bibitem{Chan:1996rd}
Kevin C.~K. Chan.
\newblock {Modifications of the BTZ black hole by a dilaton / scalar}.
\newblock {\em Phys. Rev. D}, 55:3564--3574, 1997.

\bibitem{Ayon-Beato:2004nzi}
Eloy Ayon-Beato, Cristian Martinez, and Jorge Zanelli.
\newblock {Stealth scalar field overflying a (2+1) black hole}.
\newblock {\em Gen. Rel. Grav.}, 38:145--152, 2006.

\bibitem{Banados:2005hm}
Maximo Banados and Stefan Theisen.
\newblock {Scale invariant hairy black holes}.
\newblock {\em Phys. Rev. D}, 72:064019, 2005.

\bibitem{Correa:2010hf}
Francisco Correa, Cristian Martinez, and Ricardo Troncoso.
\newblock {Scalar solitons and the microscopic entropy of hairy black holes in
  three dimensions}.
\newblock {\em JHEP}, 01:034, 2011.

\bibitem{Correa:2012rc}
Francisco Correa, An\'\i{}bal Fa\'undez, and Cristi\'an Mart\'\i{}nez.
\newblock {Rotating hairy black hole and its microscopic entropy in three
  spacetime dimensions}.
\newblock {\em Phys. Rev. D}, 87(2):027502, 2013.

\bibitem{Xu:2013nia}
Wei Xu and Liu Zhao.
\newblock {Charged black hole with a scalar hair in (2+1) dimensions}.
\newblock {\em Phys. Rev. D}, 87(12):124008, 2013.

\bibitem{Xu:2014uha}
Wei Xu, Liu Zhao, and De-Cheng Zou.
\newblock {Three dimensional rotating hairy black holes, asymptotics and
  thermodynamics}.
\newblock 6 2014.

\bibitem{Cardenas:2014kaa}
Marcela Cardenas, Oscar Fuentealba, and Cristi\'an Mart\'\i{}nez.
\newblock {Three-dimensional black holes with conformally coupled scalar and
  gauge fields}.
\newblock {\em Phys. Rev. D}, 90(12):124072, 2014.

\bibitem{Tang:2019jkn}
Zi-Yu Tang, Yen~Chin Ong, Bin Wang, and Eleftherios Papantonopoulos.
\newblock {General black hole solutions in ( 2+1 )-dimensions with a scalar
  field nonminimally coupled to gravity}.
\newblock {\em Phys. Rev. D}, 100(2):024003, 2019.

\bibitem{Dehghani:2017thu}
M.~Dehghani.
\newblock {Thermodynamics of novel charged dilatonic BTZ black holes}.
\newblock {\em Phys. Lett. B}, 773:105--111, 2017.

\bibitem{Dehghani:2017zkm}
M.~Dehghani.
\newblock {Thermodynamics of (2+1)-dimensional black holes in
  Einstein-Maxwell-dilaton gravity}.
\newblock {\em Phys. Rev. D}, 96(4):044014, 2017.

\bibitem{Bueno:2021krl}
Pablo Bueno, Pablo~A. Cano, Javier Moreno, and Guido van~der Velde.
\newblock {Regular black holes in three dimensions}.
\newblock {\em Phys. Rev. D}, 104(2):L021501, 2021.

\bibitem{Ahn:2015uza}
Byoungjoon Ahn, Seungjoon Hyun, Sang-A Park, and Sang-Heon Yi.
\newblock {Scaling symmetry and scalar hairy rotating AdS$_3$ black holes}.
\newblock {\em Phys. Rev. D}, 93(2):024041, 2016.

\bibitem{Karakasis:2022fep}
Thanasis Karakasis, Eleftherios Papantonopoulos, Zi-Yu Tang, and Bin Wang.
\newblock {Rotating (2+1)-dimensional black holes in Einstein-Maxwell-dilaton
  theory}.
\newblock {\em Phys. Rev. D}, 107(2):024043, 2023.

\bibitem{Zou:2014gla}
De-Cheng Zou, Yunqi Liu, Bin Wang, and Wei Xu.
\newblock {Thermodynamics of rotating black holes with scalar hair in three
  dimensions}.
\newblock {\em Phys. Rev. D}, 90(10):104035, 2014.

\bibitem{Sadeghi:2013gmf}
J.~Sadeghi, B.~Pourhassan, and H.~Farahani.
\newblock {Rotating charged hairy black hole in (2+1) dimensions and particle
  acceleration}.
\newblock {\em Commun. Theor. Phys.}, 62(3):358--362, 2014.

\bibitem{Zhao:2013isa}
Liu Zhao, Wei Xu, and Bin Zhu.
\newblock {Novel rotating hairy black hole in (2+1)-dimensions}.
\newblock {\em Commun. Theor. Phys.}, 61(4):475--481, 2014.

\bibitem{Bravo-Gaete:2014haa}
Moises Bravo-Gaete and Mokhtar Hassaine.
\newblock {Thermodynamics of a BTZ black hole solution with an Horndeski
  source}.
\newblock {\em Phys. Rev. D}, 90(2):024008, 2014.

\bibitem{Baake:2020tgk}
Olaf Baake, Moises~F. Bravo~Gaete, and Mokhtar Hassaine.
\newblock {Spinning black holes for generalized scalar tensor theories in three
  dimensions}.
\newblock {\em Phys. Rev. D}, 102(2):024088, 2020.

\bibitem{Bravo-Gaete:2020ftn}
Moises Bravo-Gaete and Mar\'\i{}a~Montserrat Ju\'arez-Aubry.
\newblock {Thermodynamics and Cardy-like formula for nonminimally dressed,
  charged Lifshitz black holes in new massive gravity}.
\newblock {\em Class. Quant. Grav.}, 37(7):075016, 2020.

\bibitem{Karakasis:2023ljt}
Thanasis Karakasis, George Koutsoumbas, and Eleftherios Papantonopoulos.
\newblock {Black Holes with Scalar Hair in Three Dimensions}.
\newblock 5 2023.

\bibitem{Ruffini:1971bza}
Remo Ruffini and John~A. Wheeler.
\newblock {Introducing the black hole}.
\newblock {\em Phys. Today}, 24(1):30, 1971.

\bibitem{Bekenstein:1971hc}
Jacob~D. Bekenstein.
\newblock {Nonexistence of baryon number for static black holes}.
\newblock {\em Phys. Rev. D}, 5:1239--1246, 1972.

\bibitem{Sudarsky:1995zg}
D.~Sudarsky.
\newblock {A Simple proof of a no hair theorem in Einstein Higgs theory,}.
\newblock {\em Class. Quant. Grav.}, 12:579--584, 1995.

\bibitem{Heusler:1992ss}
M.~Heusler.
\newblock {A No hair theorem for selfgravitating nonlinear sigma models}.
\newblock {\em J. Math. Phys.}, 33:3497--3502, 1992.

\bibitem{Herdeiro:2015waa}
Carlos A.~R. Herdeiro and Eugen Radu.
\newblock {Asymptotically flat black holes with scalar hair: a review}.
\newblock {\em Int. J. Mod. Phys. D}, 24(09):1542014, 2015.

\bibitem{Israel:1967wq}
Werner Israel.
\newblock {Event horizons in static vacuum space-times}.
\newblock {\em Phys. Rev.}, 164:1776--1779, 1967.

\bibitem{Wald:1971iw}
Robert~M. Wald.
\newblock {Final states of gravitational collapse}.
\newblock {\em Phys. Rev. Lett.}, 26:1653--1655, 1971.

\bibitem{Carter:1971zc}
B.~Carter.
\newblock {Axisymmetric Black Hole Has Only Two Degrees of Freedom}.
\newblock {\em Phys. Rev. Lett.}, 26:331--333, 1971.

\bibitem{Robinson:1975bv}
D.~C. Robinson.
\newblock {Uniqueness of the Kerr black hole}.
\newblock {\em Phys. Rev. Lett.}, 34:905--906, 1975.

\bibitem{Mazur:1982db}
P.~O. Mazur.
\newblock {PROOF OF UNIQUENESS OF THE KERR-NEWMAN BLACK HOLE SOLUTION}.
\newblock {\em J. Phys. A}, 15:3173--3180, 1982.

\bibitem{Mazur:1984wz}
P.~O. Mazur.
\newblock {A Global Identity for Nonlinear $\sigma$ Models}.
\newblock {\em Phys. Lett. A}, 100:341, 1984.

\bibitem{Teitelboim:1972qx}
C.~Teitelboim.
\newblock {Nonmeasurability of the quantum numbers of a black hole}.
\newblock {\em Phys. Rev. D}, 5:2941--2954, 1972.

\bibitem{Volkov:1990sva}
M.~S. Volkov and D.~V. Galtsov.
\newblock {Black holes in Einstein Yang-Mills theory. (In Russian)}.
\newblock {\em Sov. J. Nucl. Phys.}, 51:747--753, 1990.

\bibitem{Bizon:1990sr}
P.~Bizon.
\newblock {Colored black holes}.
\newblock {\em Phys. Rev. Lett.}, 64:2844--2847, 1990.

\bibitem{Volkov:1989fi}
M.~S. Volkov and D.~V. Galtsov.
\newblock {NonAbelian Einstein Yang-Mills black holes}.
\newblock {\em JETP Lett.}, 50:346--350, 1989.

\bibitem{Greene:1992fw}
Brian~R. Greene, Samir~D. Mathur, and Christopher~M. O'Neill.
\newblock {Eluding the no hair conjecture: Black holes in spontaneously broken
  gauge theories}.
\newblock {\em Phys. Rev. D}, 47:2242--2259, 1993.

\bibitem{Kanti:1995vq}
P.~Kanti, N.~E. Mavromatos, J.~Rizos, K.~Tamvakis, and E.~Winstanley.
\newblock {Dilatonic black holes in higher curvature string gravity}.
\newblock {\em Phys. Rev. D}, 54:5049--5058, 1996.

\bibitem{Luckock:1986tr}
Hugh Luckock and Ian Moss.
\newblock {BLACK HOLES HAVE SKYRMION HAIR}.
\newblock {\em Phys. Lett. B}, 176:341--345, 1986.

\bibitem{Droz:1991cx}
Serge Droz, Markus Heusler, and Norbert Straumann.
\newblock {New black hole solutions with hair}.
\newblock {\em Phys. Lett. B}, 268:371--376, 1991.

\bibitem{Ovalle:2020kpd}
J.~Ovalle, R.~Casadio, E.~Contreras, and A.~Sotomayor.
\newblock {Hairy black holes by gravitational decoupling}.
\newblock {\em Phys. Dark Univ.}, 31:100744, 2021.

\bibitem{Mahapatra:2022xea}
Subhash Mahapatra and Indrani Banerjee.
\newblock {Rotating hairy black holes and thermodynamics from gravitational
  decoupling}.
\newblock {\em Phys. Dark Univ.}, 39:101172, 2023.

\bibitem{Torii:1998ir}
Takashi Torii, Kengo Maeda, and Makoto Narita.
\newblock {No scalar hair conjecture in asymptotic de Sitter space-time}.
\newblock {\em Phys. Rev. D}, 59:064027, 1999.

\bibitem{Torii:2001pg}
Takashi Torii, Kengo Maeda, and Makoto Narita.
\newblock {Scalar hair on the black hole in asymptotically anti-de Sitter
  space-time}.
\newblock {\em Phys. Rev. D}, 64:044007, 2001.

\bibitem{Winstanley:2002jt}
Elizabeth Winstanley.
\newblock {On the existence of conformally coupled scalar field hair for black
  holes in (anti-)de Sitter space}.
\newblock {\em Found. Phys.}, 33:111--143, 2003.

\bibitem{Martinez:2004nb}
Cristian Martinez, Ricardo Troncoso, and Jorge Zanelli.
\newblock {Exact black hole solution with a minimally coupled scalar field}.
\newblock {\em Phys. Rev. D}, 70:084035, 2004.

\bibitem{Martinez:2006an}
Cristian Martinez and Ricardo Troncoso.
\newblock {Electrically charged black hole with scalar hair}.
\newblock {\em Phys. Rev. D}, 74:064007, 2006.

\bibitem{Hertog:2004dr}
Thomas Hertog and Kengo Maeda.
\newblock {Black holes with scalar hair and asymptotics in N = 8 supergravity}.
\newblock {\em JHEP}, 07:051, 2004.

\bibitem{Henneaux:2006hk}
Marc Henneaux, Cristian Martinez, Ricardo Troncoso, and Jorge Zanelli.
\newblock {Asymptotic behavior and Hamiltonian analysis of anti-de Sitter
  gravity coupled to scalar fields}.
\newblock {\em Annals Phys.}, 322:824--848, 2007.

\bibitem{Herdeiro:2014goa}
Carlos A.~R. Herdeiro and Eugen Radu.
\newblock {Kerr black holes with scalar hair}.
\newblock {\em Phys. Rev. Lett.}, 112:221101, 2014.

\bibitem{Herdeiro:2018wub}
Carlos A.~R. Herdeiro, Eugen Radu, Nicolas Sanchis-Gual, and Jos\'e~A. Font.
\newblock {Spontaneous Scalarization of Charged Black Holes}.
\newblock {\em Phys. Rev. Lett.}, 121(10):101102, 2018.

\bibitem{Hertog:2006rr}
Thomas Hertog.
\newblock {Towards a Novel no-hair Theorem for Black Holes}.
\newblock {\em Phys. Rev. D}, 74:084008, 2006.

\bibitem{Kolyvaris:2010yyf}
Theodoros Kolyvaris, George Koutsoumbas, Eleftherios Papantonopoulos, and
  George Siopsis.
\newblock {A New Class of Exact Hairy Black Hole Solutions}.
\newblock {\em Gen. Rel. Grav.}, 43:163--180, 2011.

\bibitem{Gonzalez:2013aca}
P.~A. Gonz\'alez, Eleftherios Papantonopoulos, Joel Saavedra, and Yerko
  V\'asquez.
\newblock {Four-Dimensional Asymptotically AdS Black Holes with Scalar Hair}.
\newblock {\em JHEP}, 12:021, 2013.

\bibitem{Dias:2011tj}
Oscar J.~C. Dias, Pau Figueras, Shiraz Minwalla, Prahar Mitra, Ricardo
  Monteiro, and Jorge~E. Santos.
\newblock {Hairy black holes and solitons in global $AdS_5$}.
\newblock {\em JHEP}, 08:117, 2012.

\bibitem{Bhattacharyya:2010yg}
Sayantani Bhattacharyya, Shiraz Minwalla, and Kyriakos Papadodimas.
\newblock {Small Hairy Black Holes in $AdS_5 x S^5$}.
\newblock {\em JHEP}, 11:035, 2011.

\bibitem{Dias:2011at}
Oscar J.~C. Dias, Gary~T. Horowitz, and Jorge~E. Santos.
\newblock {Black holes with only one Killing field}.
\newblock {\em JHEP}, 07:115, 2011.

\bibitem{Anabalon:2012ih}
Andres Anabalon and Julio Oliva.
\newblock {Exact Hairy Black Holes and their Modification to the Universal Law
  of Gravitation}.
\newblock {\em Phys. Rev. D}, 86:107501, 2012.

\bibitem{Kolyvaris:2011fk}
Theodoros Kolyvaris, George Koutsoumbas, Eleftherios Papantonopoulos, and
  George Siopsis.
\newblock {Scalar Hair from a Derivative Coupling of a Scalar Field to the
  Einstein Tensor}.
\newblock {\em Class. Quant. Grav.}, 29:205011, 2012.

\bibitem{Kolyvaris:2013zfa}
Theodoros Kolyvaris, George Koutsoumbas, Eleftherios Papantonopoulos, and
  George Siopsis.
\newblock {Phase Transition to a Hairy Black Hole in Asymptotically Flat
  Spacetime}.
\newblock {\em JHEP}, 11:133, 2013.

\bibitem{Ballon-Bayona:2020xls}
Alfonso Ballon-Bayona, Henrique Boschi-Filho, Eduardo~Folco Capossoli, and
  Diego~M. Rodrigues.
\newblock {Criticality from Einstein-Maxwell-dilaton holography at finite
  temperature and density}.
\newblock {\em Phys. Rev. D}, 102(12):126003, 2020.

\bibitem{Guo:2021ere}
Guangzhou Guo, Peng Wang, Houwen Wu, and Haitang Yang.
\newblock {Thermodynamics and phase structure of an Einstein-Maxwell-scalar
  model in extended phase space}.
\newblock {\em Phys. Rev. D}, 105(6):064069, 2022.

\bibitem{Anabalon:2013qua}
Andres Anabalon, Dumitru Astefanesei, and Robert Mann.
\newblock {Exact asymptotically flat charged hairy black holes with a dilaton
  potential}.
\newblock {\em JHEP}, 10:184, 2013.

\bibitem{Astefanesei:2019ehu}
Dumitru Astefanesei, Robert~B. Mann, and Ra\'ul Rojas.
\newblock {Hairy Black Hole Chemistry}.
\newblock {\em JHEP}, 11:043, 2019.

\bibitem{Priyadarshinee:2021rch}
Supragyan Priyadarshinee, Subhash Mahapatra, and Indrani Banerjee.
\newblock {Analytic topological hairy dyonic black holes and thermodynamics}.
\newblock {\em Phys. Rev. D}, 104(8):084023, 2021.

\bibitem{Mahapatra:2020wym}
Subhash Mahapatra, Supragyan Priyadarshinee, Gosala~Narasimha Reddy, and
  Bhaskar Shukla.
\newblock {Exact topological charged hairy black holes in AdS Space in
  $D$-dimensions}.
\newblock {\em Phys. Rev. D}, 102(2):024042, 2020.

\bibitem{Guo:2023mda}
Guangzhou Guo, Peng Wang, Houwen Wu, and Haitang Yang.
\newblock {Scalarized Kerr-Newman black holes}.
\newblock {\em JHEP}, 10:076, 2023.

\bibitem{Kiorpelidi:2023jjw}
Stella Kiorpelidi, Thanasis Karakasis, George Koutsoumbas, and Eleftherios
  Papantonopoulos.
\newblock {Scalarization of the Reissner-Nordsr\"om black hole with higher
  gauge field corrections}.
\newblock 11 2023.

\bibitem{Theodosopoulos:2023ice}
Dionysios~P. Theodosopoulos, Thanasis Karakasis, George Koutsoumbas, and
  Eleftherios Papantonopoulos.
\newblock {Motion of particles around a magnetically charged Euler-Heisenberg
  black hole with scalar hair and the Event Horizon Telescope}.
\newblock 11 2023.

\bibitem{Dehghani:2022aae}
Mohsen Dehghani.
\newblock {Three-dimensional black holes with scalar hair coupled to a
  Maxwell-like electrodynamics}.
\newblock {\em Mod. Phys. Lett. A}, 37(30):2250205, 2022.

\bibitem{Dehghani:2018hpb}
M.~Dehghani.
\newblock {Nonlinearly charged three-dimensional black holes in the
  Einstein-dilaton gravity theory}.
\newblock {\em Eur. Phys. J. Plus}, 133(11):474, 2018.

\bibitem{Dehghani:2019dab}
M.~Dehghani.
\newblock {Three-dimensional scalar-tensor black holes with conformally
  invariant electrodynamics}.
\newblock {\em Phys. Rev. D}, 100(8):084019, 2019.

\bibitem{Hendi:2017mgb}
S.~H. Hendi, B.~Eslam~Panah, S.~Panahiyan, and A.~Sheykhi.
\newblock {Dilatonic BTZ black holes with power-law field}.
\newblock {\em Phys. Lett. B}, 767:214--225, 2017.

\bibitem{Priyadarshinee:2023cmi}
Supragyan Priyadarshinee and Subhash Mahapatra.
\newblock {Analytic three-dimensional primary hair charged black holes and
  thermodynamics}.
\newblock {\em Phys. Rev. D}, 108(4):044017, 2023.

\bibitem{Dudal:2017max}
David Dudal and Subhash Mahapatra.
\newblock {Thermal entropy of a quark-antiquark pair above and below
  deconfinement from a dynamical holographic QCD model}.
\newblock {\em Phys. Rev. D}, 96(12):126010, 2017.

\bibitem{Dudal:2021jav}
D.~Dudal, A.~Hajilou, and S.~Mahapatra.
\newblock {A quenched 2-flavour
  Einstein\textendash{}Maxwell\textendash{}Dilaton gauge-gravity model}.
\newblock {\em Eur. Phys. J. A}, 57(4):142, 2021.

\bibitem{Bohra:2020qom}
Hardik Bohra, David Dudal, Ali Hajilou, and Subhash Mahapatra.
\newblock {Chiral transition in the probe approximation from an
  Einstein-Maxwell-dilaton gravity model}.
\newblock {\em Phys. Rev. D}, 103(8):086021, 2021.

\bibitem{Mahapatra:2018gig}
Subhash Mahapatra and Pratim Roy.
\newblock {On the time dependence of holographic complexity in a dynamical
  Einstein-dilaton model}.
\newblock {\em JHEP}, 11:138, 2018.

\bibitem{Bohra:2019ebj}
Hardik Bohra, David Dudal, Ali Hajilou, and Subhash Mahapatra.
\newblock {Anisotropic string tensions and inversely magnetic catalyzed
  deconfinement from a dynamical AdS/QCD model}.
\newblock {\em Phys. Lett. B}, 801:135184, 2020.

\bibitem{He:2013qq}
Song He, Shang-Yu Wu, Yi~Yang, and Pei-Hung Yuan.
\newblock {Phase Structure in a Dynamical Soft-Wall Holographic QCD Model}.
\newblock {\em JHEP}, 04:093, 2013.

\bibitem{Arefeva:2018hyo}
Irina Aref'eva and Kristina Rannu.
\newblock {Holographic Anisotropic Background with Confinement-Deconfinement
  Phase Transition}.
\newblock {\em JHEP}, 05:206, 2018.

\bibitem{Arefeva:2022avn}
Irina~Ya. Aref'eva, Alexey Ermakov, Kristina Rannu, and Pavel Slepov.
\newblock {Holographic model for light quarks in anisotropic hot dense QGP with
  external magnetic field}.
\newblock {\em Eur. Phys. J. C}, 83(1):79, 2023.

\bibitem{Arefeva:2020byn}
Irina~Ya. Aref'eva, Kristina Rannu, and Pavel Slepov.
\newblock {Holographic anisotropic model for light quarks with
  confinement-deconfinement phase transition}.
\newblock {\em JHEP}, 06:090, 2021.

\bibitem{Alanen:2009xs}
J.~Alanen, K.~Kajantie, and V.~Suur-Uski.
\newblock {A gauge/gravity duality model for gauge theory thermodynamics}.
\newblock {\em Phys. Rev. D}, 80:126008, 2009.

\bibitem{Gubser:2000nd}
Steven~S. Gubser.
\newblock {Curvature singularities: The Good, the bad, and the naked}.
\newblock {\em Adv. Theor. Math. Phys.}, 4:679--745, 2000.

\bibitem{Jain:2022hxl}
Parul Jain, Siddhi~Swarupa Jena, and Subhash Mahapatra.
\newblock {Holographic confining-deconfining gauge theories and entanglement
  measures with a magnetic field}.
\newblock {\em Phys. Rev. D}, 107(8):086016, 2023.

\bibitem{Shukla:2023pbp}
Bhaskar Shukla, David Dudal, and Subhash Mahapatra.
\newblock {Anisotropic and frame dependent chaos of suspended strings from a
  dynamical holographic QCD model with magnetic field}.
\newblock {\em JHEP}, 06:178, 2023.

\bibitem{Jena:2022nzw}
Siddhi~Swarupa Jena, Bhaskar Shukla, David Dudal, and Subhash Mahapatra.
\newblock {Entropic force and real-time dynamics of holographic quarkonium in a
  magnetic field}.
\newblock {\em Phys. Rev. D}, 105(8):086011, 2022.

\bibitem{Breitenlohner:1982jf}
Peter Breitenlohner and Daniel~Z. Freedman.
\newblock {Stability in Gauged Extended Supergravity}.
\newblock {\em Annals Phys.}, 144:249, 1982.

\end{thebibliography}
\bibliographystyle{unsrt}


\end{document}